\documentclass[a4paper,10pt, twocolumn]{article}

\usepackage{graphicx}
\usepackage[hyphens]{url}
\usepackage[hidelinks]{hyperref}
\usepackage{multirow}
\usepackage{tabularx}
\usepackage{cite}
\usepackage{booktabs}
\usepackage{framed}
\usepackage{arydshln}
\usepackage{color}
\usepackage{amssymb}
\usepackage{fdsymbol}
\usepackage{floatrow}
\usepackage{graphicx}
\usepackage{marvosym}
\usepackage{wasysym}
\usepackage{subcaption}
\usepackage{rotating}
\usepackage{tablefootnote}
\usepackage{eurosym}
\usepackage{appendix}
\usepackage{authblk}
\usepackage{caption}
\usepackage{longtable}
\usepackage{fancyhdr}

\DeclareUrlCommand{\tturl}{\urlstyle{tt}}
\DeclareUrlCommand{\bftturl}{}
\newcommand{\etAl}{et\,al. }
\newcommand{\eg}{e.\,g., }
\newcommand{\ie}{i.\,e., }

\newfloatcommand{capbtabbox}{table}[][\FBwidth]

\def\BibTeX{{\rm B\kern-.05em{\sc i\kern-.025em b}\kern-.08em
    T\kern-.1667em\lower.7ex\hbox{E}\kern-.125emX}}
\begin{document}

\title{\Large \textit{The Unwanted Sharing Economy:}\\
An Analysis of Cookie Syncing and User
Transparency under GDPR}

\author[1, 2]{Tobias Urban}
\author[2]{Dennis Tatang}
\author[2]{Martin Degeling}
\author[2]{Thorsten Holz}
\author[1]{Norbert Pohlmann}
\affil[1]{Institute for Internet Security, Germany \protect \\ \{lastname\}@internet-sicherheit.de} 
\affil[2]{Ruhr-University Bochum, Germany \protect \\ \{firstname.lastname\}@rub.de}
\date{}

\maketitle

\thispagestyle{empty}

\begin{abstract}
The European General Data Protection Regulation (GDPR), which went into effect in May 2018, leads to important changes in this area: companies are now required to ask for users' consent before collecting and sharing personal data and by law users now have the right to gain access to the personal information collected about them.

In this paper, we study and evaluate the effect of the GDPR on the online advertising ecosystem.
In a first step, we measure the impact of the legislation on the connections (regarding \emph{cookie syncing}) between third-parties and show that the general structure how the entities are arranged is not affected by the GDPR. However, we find that the new regulation has a statistically significant impact on the number of connections, which shrinks by around 40\%.
Furthermore, we analyze the right to data portability by evaluating the \emph{subject access right} process of popular companies in this ecosystem and observe differences between the processes implemented by the companies and how they interpret the new legislation.
We exercised our \emph{right of access} under GDPR with 36 companies that had tracked us online. Although 32 companies (89\%) we inquired replied within the period defined by law, only 21 (58\%) finished the process by the deadline set in the GDPR.
Our work has implications regarding the implementation of privacy law as well as what online tracking companies should do to be more compliant with the new regulation.
\end{abstract}

\pagestyle{fancy}
\fancyhf{}
\fancyhead[L]{\nouppercase{\rightmark}}
\fancyfoot[C]{\thepage}

\section{Introduction}
\label{sec:intro}
Today, online advertising is the most important source of income for many websites, apps, and services online. To individually target website visitors with ads, tracking services gather personal data without users' explicit consent~\cite{truste2017}. 
Whole business models formed around collecting, sharing, and using of such personal data (\eg personalized ads, analytic services~\cite{acar2014web, Englehardt2016, Starov.2017}, or malicious exfiltration of such data~\cite{Urban.2018}). This kind of data is seen as an economic asset of a company~\cite{schwab2011personal}.
As a result, imbalance of power between data processors (service providers) and data subjects (users) increased in the last years.
Often users do not know about collection, usage, and are very often unaware of the consequences~\cite{CONSENT2017}.
They also had limited options if they wanted to regain control of their data or even try to understand what information is collected or inferred about them.

To address some of these problems, the European General Data Protection Regulation (GDPR)  introduced significant changes that affect how personal data can be collected and shared. The new legislation went into effect on May 25, 2018. One of its goals is to allow users to regain control of the immaterial wealth of their personal data by introducing additional possibilities like the right to request a copy of their data, the right to erasure, and the need for services to explicitly ask for consent before collecting or sharing personal information~\cite{gdpr2016}.
Compliance with the GDPR rules is required for any company that offers services in the European Union---no matter where their headquarter is located. 
Furthermore, companies that process personal data have to demonstrate that they meet the GDPR criteria (\eg encryption and pseudonymization of personal data). Otherwise, they risk a fine of up to 4\% of the annual worldwide turnover of the preceding financial year.

Previous studies have shown that tracking and ID syncing happen in practice \cite{Englehardt2016,acar2014web}. Other works found that the GDPR affected the adoption of privacy policies and cookie notices \cite{Degeling.2018} and that it also led to a reduction of tracking \cite{WhoTracksMe2018}. The data portability right and the effects of the GDPR on cookie syncing have not been studied before.

In this work, we study the effects of the GDPR on the online advertising ecosystem and focus on the information sharing between ad services and the implementation of the new right to data portability. 
More specifically, we measure the relations of websites and third parties, and links of third parties between each other, regarding ID syncing before and after the GDPR took effect. 
To do so, we created 400 browser profiles each of which we used to visit hundreds of websites to identify ID syncing between third parties embedded in these websites.
We discovered a reduction in the number of third parties and sharing connections.
Additionally, with the top third parties in our measurement, we exercise our right of data portability and analyze the process to handle the inquiries.

Next, we found that the GDPR has (statistically) significant impact on the ID syncing phenomenon between different third parties and we found that third parties, in the ID sharing ecosystem, are often arranged like star topologies.
The most important tool for online services to inform users about their data practices are their privacy policies, but our results show that not all companies take the legal obligations seriously.
Regarding the new right to data portability, we found that some companies miss to finish the process within the legal period, set up different obstacles, and provided data in various forms and detail levels.

\smallskip \noindent
To summarize, our study makes the following contributions:
\begin{itemize}
\itemsep0em 
    \item We measure implications of the GDPR regarding the use of third-party services by websites and analyze the shift of relations between these third parties in terms of ID syncing.
    To do so, we construct an undirected graph that describes the relations between third parties and websites.
    Based on empirical measurements over a period of six months, we show that the general structure of relations is not affected, but the amount of ID sharing is reduced, by over 40\%.
    \item Furthermore, we requested access to our personal data from 39 companies and analyze the success of the \emph{subject access request}s regarding, \eg workload or timing.
    We found that 58\% of the companies did not finish the process within the legal period and that the workload to get access varies heavily depending on the inquired company. 
    \item Finally, we examine the \emph{subject access request} process of each company in detail and report on obstacles (\eg affidavits), data provided by the companies (\eg clickstream data), and information provided by companies complementary to data provided in their privacy policy (\eg specific partners who got access to our data). 
\end{itemize}

The rest of the paper is organized as follows.
We give background information most relevant for our work in Section~\ref{sec:background}  and describe our measuring approach in Section~\ref{sec:metodology}.
we determined the most prominent companies in our dataset (Section~\ref{sec:analysiscorpus} and evaluated the privacy statements regarding their compliance to the GDPR (see Section~\ref{sec:policies}).

\section{Background}
\label{sec:background}
The following sections give an overview of the GDPR's relevant regulations as well as the technical background of tracking and sharing in the online advertisement ecosystem.

\subsection{General Data Protection Regulation}
\label{sec:gdpr}

The General Data Protection Regulation (GDPR or Regulation 2016\slash 679)~\cite{gdpr2016} is a regulatory initiative by the European Union (EU) to harmonize data protection laws between its member states. After a transition period of two years, it was finally put into effect on May 25, 2018.
The GDPR specifies under which circumstances personal data may be processed and includes several rights of data subjects and obligations for those processing personal data of EU-citizens. Online advertising companies need to specify, for example in their privacy policy, for what purpose they collect and share data.
We leverage these new transparency requirements and user rights (of which the right to retrieve a copy was newly introduced) to shed light on some of the practices of online advertising services. 

GDPR articles 13 to 15 specify what information a data controller needs to provide in a privacy policy or upon request. 
This includes contact data, information about the purpose of data processing, categories of what data is used, with whom data is shared or how it was obtained, and whether it is used for profiling.
In addition, article 20 specifies a user's rights to receive a copy of data processed about her.
For any information request, including those to data access, the GDPR specifies that they must be answered within one month (Art. 12, No. 2), but can be extended by two months.
If the data controller needs more time, they must state that within a month and give an explanation("\textit{That period may be extended by two further months where necessary, taking into account the complexity and number of the requests}").

The right to retrieve a copy of data stored about a data subject is described in Art. 20. According to recital 68 of the GDPR (recitals describe the reasoning behind regulations), the \emph{right to data portability} is meant to support an individual in gaining control over one's personal data by allowing access to the data stored about him or her ``\textit{in a structured, commonly used, machine-readable and interoperable format}''.

Also relevant to our study are new regulations on profiling which is defined in Art. 22 as:
\begin{quote}
    any form of \textbf{automated processing [...] to analyze or predict} aspects concerning that natural person's performance at work, economic situation, health, \textbf{personal preferences, interests}, reliability, behavior, \textbf{location} or movements;
\end{quote}

Any company that infers information like interests about an individual for advertising purposes performs profiling and needs to disclose this. However, they are not necessarily bound to the additional requirements for profiling mentioned in the respective Article 22, \eg to enable human intervention, as profiling for advertising purposes most likely does not have any legal or significant effects.

Some tracking companies claim that the data they use is not personal information because it is pseudonymized (see Section~\ref{sec:policies}).
If this was true, it would free them from any data protection related obligations.
But the Article 29 Working Group, a committee of European data protection officials, has made clear in 2010 that storing and accessing a cookie on a users device is indeed processing of personal data since it ``\textit{enable[s] data subjects to be 'singled out', even if their real names are not known},'' and therefore requires consent~\cite{article_29_oba}. 
The document is written with respect to the previous directives, but the assessment has been confirmed by court rulings that, \eg found IP addresses, sometimes also considered pseudonyms, to be personal data.
Relevant for our study is the clarification that ad network providers, and not those that embed the third-party scripts on their websites, are responsible for the data processing.
It is argued that, since advertisers rent the space on publisher websites set cookies linked to their hosts, they are responsible for the data processing.

Organizations representing the online advertising industry interpret some of the legislation differently. The Interactive Advertising Bureau (IAB) Europe has published a working paper for data subject requests in April 2018~\cite{iabeurope}. They focus on two provisions that limit the data processor's obligations to answer requests from individuals. First, Article 12 states that a data controller does not need to act on requests if they ``\textit{demonstrat[e] that [they are] not in a position to identify the data subject.}'' Second, Recital 64 of the GDPR states that data controllers should ``\textit{use reasonable measures to verify the identity of a data subject who requests access}'' to make sure they do not give away personal information to someone that is just impersonating someone else to get access to information.
The argument is that if the data is used in a pseudonymous fashion, \ie a cookie ID, data subjects need to prove that the information connected to that ID is actually about them.
Still, the guidelines argue that services should make individual decisions as to how they respond to those requests, and should, \eg not provide clickstream data but information about the segments an ID is assigned to.

\subsection{Advertising Economy}
\label{sec:ad_economy}

Displaying ads is the most popular way to fund online services. 
In 2016, the online advertising industry generated \$72.5 billion Dollars in the United States~\cite{IAB2016}, while it generated  \EUR{41.8} billion Euros in Europe in 2017~\cite{IAB2017}.
The modern online advertisement ecosystem is quite complex and is, in a nutshell, composted out of three basic entities which are described in the following~\cite{Yuan2014}. 

On the one end, there are publishers and website owners that use  \textit{supply-side platforms} (SSP) to sell ad space on their sites.
On the other end, the \textit{demand-side platform} (DSP) is used by marketing companies to organize advertising campaigns, across a range of publishers (SSPs). 
To do so, they not necessarily have to select a specific publisher they want to work with, but can define target users based on different criteria (\eg geolocation, categories of websites visited, or personal preferences). 
A \textit{data management platform} (DMP) captures and evaluates user data to organize and optimize digital ad campaigns. They can be used to merge data sets and user information from different sources and use that data to automate campaigns on DSPs.
To do so, a DMP often collects IDs of different systems and merges data with those from other sources to target ad campaigns to a specific audience based on high-level information like interests. 

User tracking and profiling are important parts of a website's or other application's (\eg mobile applications) business model ~\cite{Starov.2017, acar2014web, Englehardt2016}.
Profiles are based on the users' clickstream (a list of websites a user has visited) to enable target advertising~\cite{bucklin2003model,profiles2016}.
To do so, a unique digital identifier is assigned to each user, either by the server or computed based on properties of the user's device (called \emph{device fingerprinting}~\cite{Englehardt2016}).
The most prevalent way to store such digital identifiers on a users device are \textit{HTTP cookies}.
While this is considered an undesirable privacy intrusive behavior by some, it is in practice a fundamental part of the online ad economy to perform so-called \textit{Real-time Bidding} (RTB).
RTB means that impressions and online ad space are sold in real-time on automated online marketplaces whenever a user visits a website.

Different approaches to counter user tracking have been proposed~\cite{kontaxis2015tracking, obufscation_2018} and it was shown that these approaches are often not very effective~\cite{Merzdovnik.2017}. 
One reason for this is that user tracking methods are continually evolving and sometimes get highly sophisticated (\eg using the HTML5 \texttt{canvas} tag)~\cite{Englehardt2016, Eckersley.2010}.

\begin{figure*}[tb]
    \caption{Different types of cookies: (A) a first party cookie---directly set by the visited website, (B) a third-party cookie---set by a third-party embedded in the website, and (C) a synchronized cookie---shared between two parties.}
    \label{fig:cookie_sync}
    \centering
    \includegraphics[width=.75\textwidth]{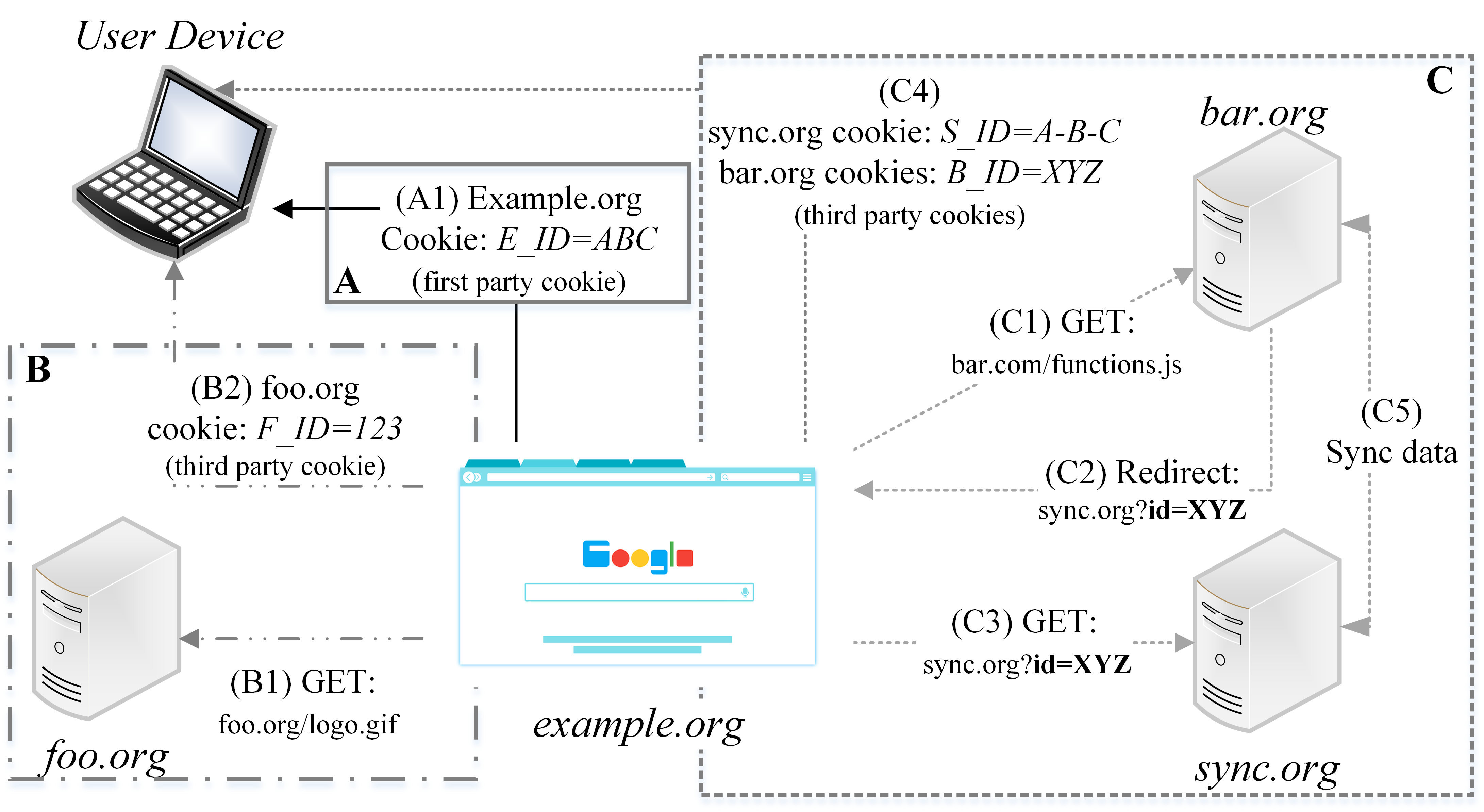}
\end{figure*}

\subsection{Cookie Synchronization}
\label{sec:cookie_syncing}

A \textit{HTTP cookie} is a piece of textual data, strictly limited in size, that can be set by a website to store data locally on a client.
In theory, cookies contain simple \texttt{name=value} pairs but in practice, they sometimes combine information through various means~\cite{gonzalez2017cookie}.
If a website visited by the user directly sets a cookie, it is called \textit{1st party cookie} (depicted in Figure~\ref{fig:cookie_sync} as (A)).
A cookie is called \textit{third-party cookie} if the visited website embeds an object from another domain and this third-party sets a cookie (see (B) in Figure~\ref{fig:cookie_sync}).

Cookies are intended to maintain a state between different HTTP sessions, \eg to remember user preferences like items stored in the shopping cart, or to log that a user has previously authenticated with the server.
Storing a unique user identifier in a cookie (\eg \texttt{uid=1234-abcd-f1}) allows a server to identify a user that is revisiting the website.
It is also common that additional information exceeding the size allowed for cookies is stored on the server side in relation to that same ID. For online advertising, this could be profile information like interest segments or geolocation.
A server can only access a cookie under the domain that set it, meaning that different third parties cannot access each others cookies. 
This prohibits data leakage or cross-domain tracking of different third parties by merely accessing the cookies (via the same \textit{Same-Origin Policy}).

\textit{Cookie syncing} is a process to bypass the Same-Origin Policy by sharing the unique identifier of a user between two third parties (see Section (C) in Figure~\ref{fig:cookie_sync}).
Cookie syncing is mostly a two-step process: (C1) A script from a third-party (\texttt{bar.org}) is loaded into a website (\texttt{example.org}). 
The request, that loads the script, is then redirected, or the script itself issues a new request to the syncing partner (\texttt{sync.org}). 
This redirected request contains the ID \texttt{bar.org} assigned to the user (\eg \texttt{sync.org?\textit{bar\_user\_id=XYZ}}).
Now \texttt{sync.org} knows, via the HTTP referrer header or additional information added to the request, that the user with \texttt{bar.org}'s ID visited \texttt{example.org}.
If \texttt{sync.org} already has a cookie (\eg from a previous visit to another website) on the client, it can pair \texttt{bar.org}'s user ID to its own.
This allows \texttt{sync.org} and \texttt{bar.org} to share data about the user over another channel. 
This mechanism also allows a tracker (\texttt{sync.org}) to track users on a large variety of websites even if these websites do \textit{not} embed the tracker but a partner of the tracker.

In most cases, the publisher of a website has little to no influence on whether or not cookie syncing is performed.  
From a privacy perspective, these techniques have been criticized as personal data is shared between several third parties without the user's knowledge or consent~\cite{Nunes2018}.

\textit{Re-targeting} is a mechanism where an advertiser ($A$) wants to target customers ($C$) with ads for products she had showed interest in on $A$'s website. 
To do so, $A$ embeds an object from an ad company ($R$), specialized in re-targeting users, onto every website $A$ runs which allows $R$ to assign an ID to $A$'s users.
$R$ syncs this ID to an ad exchange marketplace ($M$) so that M can identify if $C$ visits any website that uses $M$.
Then, if $C$ visits a website that embeds $M$, $M$ can inform $A$ that $C$ is currently visiting the website which makes it possible for $A$ to provide a targeted ad for $C$.
\section{Measurement Approach}
\label{sec:metodology}
To gain insights into information sharing between tracking companies and the impact of the GDPR on these practices, we first conducted a measurement study of cookie syncing in the browser. To better understand the sharing practices not observable through this method, we leveraged GDPR access rights and performed an in-depth study on the \emph{subject access request} (SAR) process for companies in the online ad economy.

\subsection{Measurement Framework}
\label{sec:framework}
To measure the extent of tracking and information sharing, we used the \texttt{openWPM}~\cite{Englehardt2016} platform.
For our first study, we deployed the platform on two computers at a European university. Note that a European origin of our web traffic was necessary to later be able to refer to European legislation.
We chose \emph{not} to use a scalable web service (\eg Amazon EC2) to automate our measurement since it might be easier for a website to detect such automated crawls~\cite{CloakofVisibility2016}.
For comparison, we also conducted one evaluation using US-based IP addresses (see below).

With \texttt{openWPM} we logged all HTTP request and response headers, HTTP redirects, and POST request bodies as well as various types of cookies (\eg Flash cookies).
We did not set the ``Do Not Track'' HTTP header and did allow third-party cookies.
We used some ``bot detection mitigation techniques'' (\ie scrolling randomly up and down on each visited website) to make it harder to detect our crawler.

For our analysis, we created 400 browsing profiles.
We chose the top 20 countries with the highest number of Internet users~\cite{InternetWorldStats2018} and created 20 profiles for each country.
For each of these profiles, we took the Alexa top 500 list of the corresponding country~\cite{Alexa2018} and randomly chose 100 to 400 websites to be visited.
Finally, we randomly visited three to five sub-sites from all links on the websites that were hosted under the same domain.
Each profile had its own browser profile during the analysis to make sure cookies could be separately stored for each session.
Furthermore, we randomly assigned a popular user-agent string and a standard screen resolution to each browser profile that remained constant during the analysis.

In October 2018, we performed one reference measurement with US-based IP addresses via a VPN connection to compare the results with Europe based traffic from the same time. Note that VPN services can potentially inject content (\eg ads) into the traffic that might affect the results~\cite{Khan.2018}. However, the Terms of Service did not state that this might happen nor have we found any information about content injection for this VPN service.

\subsection{Identification and Mapping of Third-Party Relations}

To analyze the sharing of \emph{personal} or \emph{digital identifiers} (IDs), we first had to filter them out.
For every visited domain, we analyzed the HTTP GET and POST requests and split the requests at characters that are typically used as delimiters (\eg '\&' or ';'). As a result, we obtained a set of ID candidates that we can store for later analysis as key-value pairs.

We identified IDs according to the following rules, which are applied to every ID candidate pair we find during the extraction phase. Note that the rules are inspired by the work of Acar \etAl\cite{acar2014web}:

\begin{itemize}
\itemsep-0.5em 
    \item Eliminate all ID candidates that were observed for multiple profiles.
        Every identifier should be unique to each user profile (\eg we eliminate the candidates $c1 = (p\_id, 1234abcd)$ and $c2 = (p\_id, 1234abcd)$ if they were observed in two different browser profiles).
    \item Eliminate ID candidates with the same key but where values differ in length.
        We expected that IDs are of consistent length (\eg the candidates $c1=(data, 3rw3)$ and $c2=(data, 70g63b5g)$ would be eliminated).
    \item Eliminate candidates whose values do not contain enough entropy (according to the Ratcliff\slash Obershelp pattern recognition algorithm~\cite{ratcliff1988pattern}) to contain an ID.
    Since we only observe a small fraction of the potential ID space, we expect that IDs to differ significantly (\eg the candidates $c1=(id, AAAC)$ and $c2=(id, AABA)$ would be eliminated).
    \item Exclude candidates whose length is too short to contain enough entropy to hold an ID.
       To provide enough entropy to differentiate between individual users, we expect an ID to have at least eight characters (\eg the candidate $c = (key, 1hgtz)$ is excluded).
\end{itemize}

To measure the syncing relations of third parties, it is necessary to identify URLs---that contain user IDs---inside a request (\eg \bftturl{foo.com}\tturl{/sync?partner=https
To do so we attempt to decode (\eg \texttt{BASE64}) and deflate (\eg \texttt{gzip}) every HTTP GET and POST argument.
Since any of these arguments might be encoded\slash inflated multiple times, as observed by Starov \etAl\cite{Starov.2017}, we repeated this process multiple times (if necessary).
We used regular expressions to parse the decoded values for URLs.
When an URL was found, we check if this URL has GET parameters that might be an ID, according to our previously described definition of an ID.

We used the \emph{WhoTracks.me} database~\cite{WhoTracksMe2018} to cluster all observed third-party websites based on the company owning the domain.
These clusters served as nodes for the construction of a graph and added two types of edges to the undirected graph to connect the nodes:  (1) directs relations (\ie a websites embeds a third-party object), and (2) syncing relations (\ie two third parties that perform cookie syncing).
Thus, we can measure (1) how many websites make use of a specific third-party, and (2) with how many other third parties IDs were synced.  

\subsection{Analysis Corpus}
\label{sec:analysiscorpus}

During the first measurement (see Section~\ref{sec:measurments}), we extracted the top 25 third parties embedded by websites as well as the top 25 third parties that engaged most in cookie syncing.
In total, we identified 36 companies (third parties) which we refer to as \emph{analysis corpus} (see Table \ref{tab:ana_corpus} in Appendix~\ref{app:corpus}).
In the remainder, if not stated otherwise, our analyses always refer to this corpus.
In three cases we were told to address our inquiry to another company (\eg subsidiary or the parent organization).
Thus, our corpus consists of 39 companies.

The complexity to measure that ecosystem is highlighted in previous work~\cite{Guha2010, barford2014}. To the best of our knowledge, there is no reliable public information on market shares in the online advertising ecosystem.
Thus, we used the third parties which are most popular in our experiment. 

The top company that we did \emph{not} include in the corpus (\ie the 26th most emended company) is embedded by 0.12\% of the visited websites and the top cookie syncing company \emph{not} included in the corpus accounts for 0.58\% of the syncing connections in the graph.
The 39 companies in the corpus account for 66\% of all cookie syncing activities (in the measurements we describe in Section~\ref{sec:framework}) while the reaming 33\% are made up of 352 companies (i.e., there is a long tail).
The companies in the corpus represent 61\% of the (directly) embedded third parties.
Due to the low share of the other companies, we did not include them in our analysis: contacting ten more companies (an increase of 19\%) would only increase the amount of covered cookie syncing by at most 5.8\% or embedded websites by at most 1.2\%. 
The actual distribution is also displayed in Figure~\ref{fig:cdf}.

\begin{figure}[tb]
    \caption{Shares of observed companies in terms of directly embedded (black) and actively syncing cookies (gray).}
    \label{fig:cdf}
    \centering
    \includegraphics[width=1\textwidth]{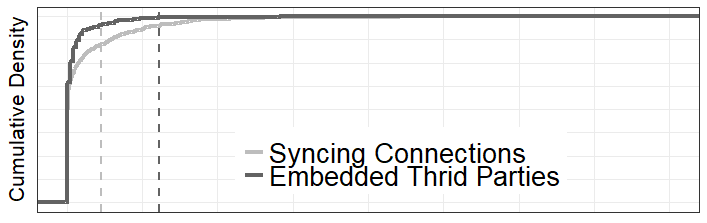}
\end{figure}

The corpus consists of six SSPs, nine DSPs, seven companies that specialized in targeted ads, four DMPs, and 13 companies whose primary business field is not directly tied to the ad economy but instead utilizes ads to finance their services (\eg \emph{RTL Group} - a Luxembourg-based digital media group, \emph{Facebook} - an online social network and media company, or \emph{Verizon} - a telecommunications company). 

While most of the companies in our corpus operate globally and run multiple offices, 82\% have their headquarters located in the United States.
The remaining 18\% are located in Europe (France, Luxembourg, the Netherlands, and the United Kingdom). 
This distribution is likely based towards US\slash EU-based companies since we run our measurements from Europe.
This distribution is likely biased towards EU-based companies and a result of our selection process from Alexa top lists of European countries. Since our goal was to measure the impact of the new European legislation we expected this bias and try to account for it in the analysis of the results.

The observed third parties are not a complete list of all companies involved in online advertising. Some third parties might only be used in specific use cases (\eg video playback), or different companies might be used based on the geolocation of the user. 
We address this limitation in further detail in Section~\ref{sec:limitaions}.

\subsection{GDPR requirements}
\label{sec:gdprmetrics}
We analyzed the privacy policies of all 39 companies described above to see whether they contain the information required by the GDPR (see Section~\ref{sec:background}). We specifically looked for information on the data sharing practices and evaluated how data subjects can exercise their rights. 
As described above, data controllers are required to inform, besides other things, about the legal basis for their data collection, categories of companies they share the data with, and how long the data is stored. 
We do not report on observations that all policies had in common but focus on the differences.
On the one hand, for example, the right to withdraw consent has been implemented through various opt-out mechanisms\cite{Degeling.2018} that all services support and are therefore not listed.
On the other hand, few services actually follow the Do-No-Track signal, although it was designed as a common consent mechanisms. Therefore, we listed statements about the latter.
We were also interested in how companies deal with the requirements regarding profiling: 
If they use profiling, they are expected to describe \emph{the logic involved} in this process, although the debate about what that should include is still ongoing~\cite{selbst_2017}.
Privacy policies should list the rights of the data subjects, \eg to object to the processing and the possibility to access the data and they should describe how these rights can be exercised. 

\subsection{Assessing the SAR Process}
\label{sec:method:sar}

One goal of the GDPR is to give users possibilities to regain control of their personal data.
In order to test to which extend users can actually exercise these rights, we contacted the companies in the corpus and looked for a preferred way of contacting them in their privacy policies.
According to Art. 13 and 14 of the GDPR, companies need to provide contact details of a responsible person (\eg the Data Privacy Officer) to handle privacy-related requests.
Most companies named a general email address to handle such requests or referenced a web form which can be used to access the data. 

To access the personal data companies stored associated with a cookie ID, we created a profile specifically for this process.
We used \texttt{openWPM} to randomly visits websites that include third parties, owned by the companies in our analysis corpus.
From these websites, all internal links (subsites) are extracted and visited in a random order.
For this analysis, we kept the session active and continued visiting websites while we requested information about the profiles.
After running the profile for approx. 72 hours, we extracted the ID key-value pairs from the cookie store and cookie syncing requests.
This \texttt{openWPM} instance was left running until the end of our analysis in order to keep the cookies active. 

When sending out inquiries, we included all cookie ID values and domains for which we observed ID syncing (with the corresponding ID key-value pairs). 
If we could add custom text to our request (via email and in some web forms), we asked four questions regarding the usage of our  data:
\begin{enumerate}
\itemsep-0.5em 
    \item What information about me\slash associated with that cookie do you store and process?
    \item Where did you get that information from? Did you get it from third parties?
    \item Do you use the data to perform profiling?
    \item With whom do you share what information and how?
\end{enumerate}
A template of the emails can be found in Appendix~\ref{app:mail}.

We used informal language (\eg we did not quote any articles from the GDPR nor did we use any legal terminology) because we wanted to assess the process when an ``average user'' wants to exercise his\slash her right to access the data.
In previous work on vulnerability notifications, a more technical and formal language was used by the authors~\cite{Li2016, Stock16}.
From our point of view, the situation is different because we wanted to exercise a right and not inform someone regarding a security problem.

An ``average user'' might have trouble to access the information we added in our mails (\eg the correct cookie value). 
However, some companies offer easy ways to access the information we included in our mails (\eg a web form that automatically reads the user ID from the browser's cookie store\cite{Sharethrough.2018}). 
Thus, we assume that a user who has privacy concerns can obtain this information and usability improvements might follow in the near future.

We conducted two rounds of inquiries. 
The first round on June 20th, approximately one month after the GDPR took effect, and the second round starting on September 21st, approx. four months after the GDP took effect.
This allows us to compare how their process changes as companies get more experienced with the GDPR in practice.

We used two newly created \texttt{GMail} accounts (one for each round of contacting) to get in touch with the companies and did \emph{not} disclose that we were conduction this survey as part of a scientific study because that could bias the responses of the companies ranging from not answering at all to giving more precise responses than they usually would.

In our measurement, we used two deadlines. 
The first deadline is the legal period defined in the GDPR, 30 days after the inquiries (July 20th and October 22nd), and the second, more relaxed deadline, is 30 \emph{business} days after our inquiries (August 1st and November 5th).
\section{Results and Evaluation}
\label{sec:evaluation}
We conducted seven measurements throughout six months. 
The first measurement was performed between 19th of May 2018 and 23rd of May 2018, just days before the GDPR went into effect, the second one after the 25th of May 2018. The other measurements were made in intervals of four to six weeks after that. The graphs were used to measure the extent of third parties and ID sharing between tracking companies (Sections~\ref{sec:partners} and~\ref{sec:measurments}). In addition, we analyzed the privacy policies of companies we requested data from (Section~\ref{sec:policies}) and evaluated the \emph{subject access requests} (SAR) of the companies (Sections~\ref{sec:sar} and~\ref{sec:process}).

\subsection{Third-Party Sharing Ecosystem}
\label{sec:measurments}

We constructed a graph for each individual measurement that represents embedded third parties and information sharing networks (see Section~\ref{sec:metodology}). The graphs are undirected, where each node represents a company and each edge represents ID syncing between companies.
Figure~\ref{fig:nodes_regression} illustrates the number of nodes per measurement. 
The x-axis represents the number of nodes and the y-axis represents the calendar weeks. 
The light gray dot on the left is the first measurement M\#1 \emph{before} the GDPR came into effect and the further darker gray (black) dots represent the corresponding other measurements (M\#2 to M\#7).
We performed two types of linear regression analysis including the measurement before the GDPR took effect (gray) and excluding (black) it.
Both trends are significant ($p$-values 0.00044 and 0.00175). Therefore, we determine that the GDPR makes a significant difference in the slope of the regression lines and thus the introduction of the GDPR resulted in a measurable decrease in embedded third parties.
As the trend progresses, the trend continues to flatten out. 
Before the GDPR enforcement, the Graph M\#1 contained 12,304 nodes, 11,738 of which are isolated. Isolated nodes have no connection to another node and represent third-party companies that are embedded into websites but do not perform cookie syncing (\eg a JavaScript library).
Overall, the number of third parties, isolated or not, decreases over the course of our study.

Figure~\ref{fig:sync_regression} shows the numbers of ID sharing connections.
Of particular interest is the reduction of syncing relations by about 50\% before and after the introduction of the GDPR, which indicates that the sharing of data is reduced --- in terms of the number of direct syncing connections.
The corresponding linear regression analysis confirms that both trends with (gray) and without the measurement before the GDPR (black) are both significant ($p$-values 0.0072 and 0.0022). This indicates that the introduction of the GDPR resulted in less direct ID syncing connections.
To assess the GDPR's impact, we compared the regressions including ($y_{pre}$, dotted, gray lines) and excluding ($y_{post}$, dashed, black lines) the pre-GDPR measurement. In both cases, the slopes are lower which indicates that the around 25th of May the drop was significantly larger, but also that the general trend is towards using fewer third parties that also sync less.
These results suggest that companies removed several third parties from their websites in order to avoid problems regarding the new legislation, an observation in line with other studies~\cite{FutureScot2018,tracking2018}.

\begin{figure*}[htb!]
    \centering
    \begin{subfigure}[c]{0.45\textwidth}
        \centering
        \includegraphics[width=0.95\textwidth]{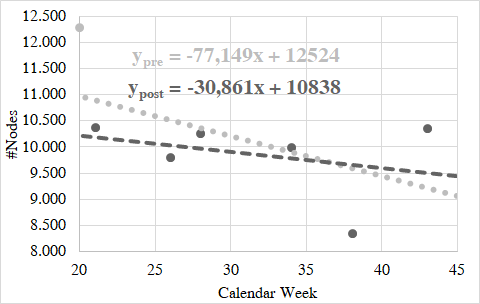}
        \subcaption{Number of nodes per measurement}
        \label{fig:nodes_regression}
    \end{subfigure}
    ~
    \begin{subfigure}[c]{0.45\textwidth}
        \centering
        \includegraphics[width=0.95\textwidth]{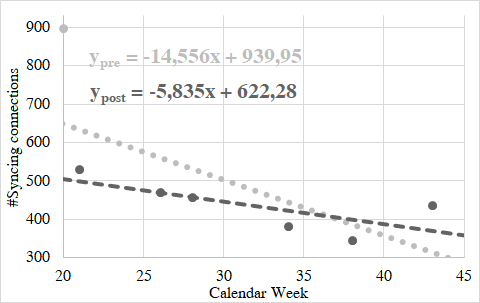}
        \subcaption{Number of syncing connections per measurement}
        \label{fig:sync_regression}
    \end{subfigure}
    \caption{Regression lines of our measurements including the pre-GDPR measurement (gray) and excluding it (black).}
\end{figure*}

Table~\ref{tab:graphs} provides an overview of the connections within the graphs. 
To measure whether the effects on the number of third parties and syncing are independent, we separated the graphs into subgraphs, so-called components. Each component represents a graph in which all nodes are connected to each other by paths.
M\#1 has 59 components, with the largest component containing 429 nodes.
The percent values reflect the reduction and always refer to the initial graph M\#1, so the number of components is reduced from M\#1 to M\#7 by maximum around 56\%.
Another difference is the size of the largest component that is reduced by up to 47\%.
However, the median component size remains stable at around two throughout all measurements. This indicates that the introduction of GDPR has no effect on the individual components.
Similarly, the algebraic connection is a measure for the number of nodes and the number of connections between the nodes within the graph. The lower the value, the fewer connections are present.
The values of the algebraic connection varies between positive 25\% and negative 60\% compared to the initial measurement.
The evaluation shows that the total number of links in the graph fluctuates, but numbers are similar comparing the first and the last measurement (-3.4\%). This shows that, although individual measurements vary, due to how the ecosystem works, the introduction of the GDPR has no direct effects on the structure of our graphs.
In comparison the reduction in the number of nodes, and therefore the number of third parties involved in tracking, follows a downward trend: Fewer third parties are present on websites (see Figure \ref{fig:nodes_regression}).

Comparing the results from computers in Europe with our reference measurement via a US-based service, we observed that the amount of cookie syncing---in terms of ID syncing connections---for website visits from the USA is about 15\% \emph{above} the amount for similar visits from the EU, in a comparable time (CW43). 

The regression analyses performed verify that the introduction of GDPR had a measurable impact on the number of nodes and links.
The number of companies is decreasing, as is the number of direct connections. The general trend is slightly downwards but stabilizes again from the last measuring point.
The general graph characteristics, however, indicate, just as the similar modularity and the similar average clustering coefficient as well as the values for the average degree, that the GDPR has not fundamentally changed the ecosystem, but resulted in a general reduction of information exchange (see also Appendix~\ref{app:graphscharacteristics}). 

As already mentioned, the results indicate that the GDPR did not change the characteristics of the ad ecosystem (\eg cookie syncing still exists) but has a significant impact on the amount of sharing using this technique.
This hints that companies are sharing with a smaller number of partners.
However, other studies show that tracking of smaller ad companies was reduced, while tracking by the market leader (\ie \emph{Google}) grew~\cite{tracking2018}. 

\begin{table}[t]
    \centering
    \caption{Overview of connected components (CP) in the measured graphs, and the shift after the GDPR took effect.}
    \label{tab:graphs}
    \resizebox{\textwidth}{!}{
        \begin{tabular}{c|rr:rr:rr}
            \toprule
            \multirow{2}{*}{\hfil CW}  & \multicolumn{6}{c}{Connectivity} \\
             & \multicolumn{2}{c:}{Components} & \multicolumn{2}{c:}{largest CP}& \multicolumn{2}{c}{algebraic conn.} \\
            \midrule        
            20 (M\#1) & 59 & ---     & 429 & ---   &  0.1187 & --- \\
            
            21 (M\#2) & 38 & $-35.59\%$ & 296 & $-31.00\%$ & 0.1494 & $+25.86\%$\\
            
            26 (M\#3) & 37 & $-37.29\%$ & 269 & $-37.30\%$ & 0.1071 & $-9.77\%$ \\     
            
            28 (M\#4) & 30 & $-49.15\%$ & 277 & $-35.43\%$ & 0.0994 & $-16.26\%$ \\ 
            
            34 (M\#5) & 37 & $-37.29\%$ & 235 & $-45.22\%$ & 0.0818 & $-31.09\%$ \\
            
            38 (M\#6) & 26 & $-55.93\%$ & 225 & $-47.55\%$ & 0.0469 & $-60.49\%$ \\
            
            43 (M\#7) & 38 & $-35.59\%$ & 268 & $-37.53\%$ & 0.1146 & $-3.45\%$ \\
            
            \bottomrule
    \end{tabular}}
\end{table}

\subsection{Connections of Third Parties}
\label{sec:partners}

As described above, we differentiated between direct partners as well as indirect partners of a third-party.
``Indirect partners'' are (recursive) partners of a direct partner which is relevant for the classification below. 
We identified three types of third parties (nodes): (1) nodes with predominately direct partners, (2) nodes with only one partner but a large number of indirect partners, and (3) partners with a rather balanced amount of direct and indirect partners.
We labeled a node ``central'' if it has four times more direct partners than indirect partners, ``outer'' if it has four times more indirect partners than direct partners, and balanced otherwise.
In our data set, we have 16 central nodes, 10 balanced nodes, five central nodes, and eight isolated nodes.
The remaining nodes in the graph (\ie not present in the analysis corpus) are mostly the outer corners in a star (see below).

We observe that the networks of cooperating third parties are arranged in star topologies (Type 1).
They have one central point which has many direct syncing connections to partners (\eg \emph{Google}), but these partners rarely sync with additional partners.
Other nodes with many indirect partners (Type 2) often have few direct partners (often just 1) which are the central point of a star.
Thus, these companies are connected to all outer corners of the star as indirect partners.
Nodes with a balanced amount of direct and indirect partners (Type 3) do not have any other special characteristics.

Over the course of our study, we could observe that the number of direct partners of most companies continuously decreased by up to 40\% (83 less direct partners). 
Five companies even got totally isolated in our measured graphs and only two companies gained direct partners.
With respect to indirect connections, we see a large fluctuation of partners that can be explained by the fact that adding one direct partner, who might be the center of another star, can lead to a significant number of additional indirect partners (sometimes hundreds of indirect partners).

Our results show that embedding one third-party into a website puts users at risk that their data gets shared with hundreds of companies.
According to the GDPR, companies do not have to disclose with which partners they share which data with, but only need to mention the categories of partners (see Section~\ref{sec:gdpr}).
This leads to the problem that users cannot verify who has received a copy of information about them and leads to the questions how service provider can ensure that data is deleted upon request.

Aside from the one dominating star, with \emph{Google} as a central point, we observe many smaller star topologies who share IDs with each other.
This is in line with our observation of the communities in the graph (see Table~\ref{tab:graphs}) and public announcements of companies to build tracking infrastructures aside from \emph{Google} or \emph{Facebook}~\cite{Nunes2018}.
The results further show that our corpus includes nodes of all types.
The classification results of the companies in our analysis corpus can be found in Appendix~\ref{app:partners}.

\subsection{Privacy Policies}
\label{sec:policies}

In addition to the measurement of data sharing of third parties in the real world, we also analyzed the privacy policies of 39 companies to check whether or not they fulfill the requirements described in Section~\ref{sec:gdprmetrics}. The most relevant details are reported in Table~\ref{tab:policies} (in Appendix~\ref{app:pp_overview}). The policies of all but three companies fulfill the minimum requirements for privacy policies. All companies offer the possibility to opt-out of their services and all but one disclose that they share some information with others, but only a few are transparent about who these third parties are and what type of information is shared. Only two of the policies disclosed and explain cookie syncing. Similarly, only eight mention that they do or do not perform profiling. One company did not update its privacy policy since 2011 and contained false claims, for example, that IP addresses are non-personal information. The lowest amount of information that we were looking for could be found in the privacy policy of \emph{Amazon}.

All policies that mention the legal basis for processing claim that they rely on individual consent when processing data, but at the same time the majority does not adhere to the (not legally binding) Do Not Track (DNT) standard, where information about whether or not users want to be tracked is conveyed in an HTTP header~\cite{donottrack_2011}. Instead, companies rely on implicit consent, meaning that as long as a data subject has not objected to a data collection by opting-out, it is assumed that users are OK with the data collection.

Differences can be found on topics specific to GDPR. For example regarding the question whether processed data that contains sensitive information (\eg about race or health). While some explicitly forbid to collect this information through their services, others acknowledge that some interest segments they provide might be health-related \eg about some beauty products. Only three companies acknowledge that they process health information, but do not discuss how this data is better protected than the rest.

\subsection{Subject Access Requests}
\label{sec:sar}

Companies also provide different means to grant data subjects access to their data. Table~\ref{tab:policies}, shown in Appendix~\ref{app:pp_overview} due to its size, lists the various forms through which users can retrieve a copy of their personal data. 
To learn more about the sharing of personal information and to fill the blank the privacy policies had, we examined how third parties adopt the new requirements of the GDPR (see Section~\ref{sec:gdpr}) and how they respond to our \emph{subject access requests} (SAR).
We contacted the companies in our analysis corpus and tried to obtain a copy of the data associated with a cookie ID to evaluate the SAR process of each company. 
As described in Section~\ref{sec:method:sar}, we used \emph{GMail} accounts to get in touch with the companies. We did so on June 20th, 2018 (round 1) and on September 21th, 2018 (round two).
In the first round, we sent out 32 emails and used six web forms to get in contact with each company.
In the second round, we sent 27 mails and used eleven web form because the contact mechanisms had slightly changed. 
As part of this process, we extracted the cookie ID values and up to five domains for which we observed ID syncing (with the ID key-value pairs) from the long-running profile in the e-mail.
The GDPR requires companies to grant users' access to their data within 30 days after their initial request. Since it does not specify whether this referrers to business or calendar days we marked two deadlines (dotted, gray lines in Figures~\ref{fig:timing} and~\ref{fig:success}). 

\subsubsection{Response Types and Timing}
We grouped responses in three types: (1) \emph{automatic} responses, (2) \emph{mixed} responses, and (3) \emph{human} responses.
A message was categorized as ``automatic'' if it was identifiable as sent automatically by a computer system (\eg a message from a ticket system stating that our request was received).
We labeled a message \emph{mixed} if the message did not directly refer to any of our questions but only included very generic information that respond to any privacy-related request.
We double checked this with mails we got in both inquiry rounds.
Messages that directly responded to our questions were labeled ``human''.
If we had any doubt, we labeled messages in favor for the companies.
Figure~\ref{fig:timing} shows amounts and type of responses we got during our analysis.
We did not count status messages from ticket systems (\eg a message stating that our ticket was updated), but only looked at those messages that contained an actual reply.

In our second round of inquiries, we received in total less responses from the companies.
This is partly because we did not have to report any broken forms to the companies which explains the fever human responses in weeks 1 and 2.
However, we observed that in our second round, companies did not follow up further questions like they did in round 1 (\eg if we asked for further clarification about data sharing). 

In round one during the first two weeks, we received by far the most responses (51), while most responses are labeled \emph{human} (56.9\%) and 25.5\% are labeled \emph{automatic}. 
While the share of response types stayed balanced, the number of responses significantly decreased although we asked follow-up questions.
In round two, we labeled 17.4\% of the responses \emph{human}, in week one and two, and 60.9\% \emph{automatic}.

In round one, we received almost exclusively responses from human correspondents one week before the deadline. That changed in round two where we got almost no responses anymore after week four (3 in total).
After the first deadline, we received mostly \emph{human} or \emph{mixed} responses. 
One company told us (in both rounds) that due to  complexity of our inquiry that they would need more time.
After the second deadline (30 business days after our inquires), we still got \emph{human} responses to our questions (ten in round one and two in round two).
The companies that did reply after the first deadline often just referenced to their privacy statement without addressing our questions directly.

\begin{figure}[ht]
    \caption{Types and timings of the received responses.}
    \label{fig:timing}
    \centering
    \includegraphics[width=1\textwidth]{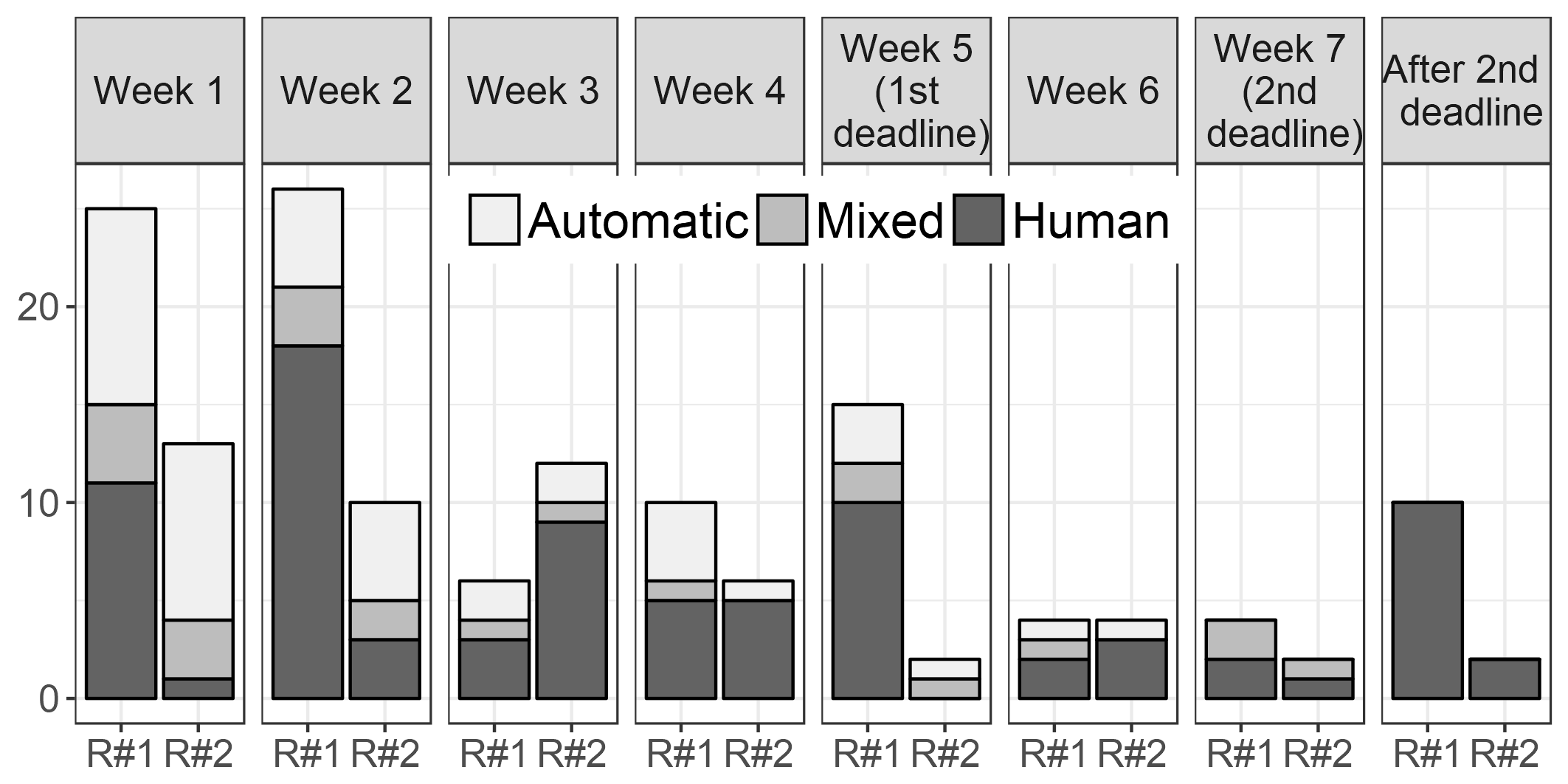}
\end{figure}

\subsubsection{Response Success}
\label{sec:success}

The amount of effort necessary to obtain access to personal data differed depending on the inquired company.
To asses the work load of the process of a company, we use a simple scoring mechanism that essentially takes four factors with different impact into account: (1) amount of emails sent to the company ($C_i$) before getting access to data associated to the digital ID ($M_{pre}$), (2) amount of emails sent after getting access to the data ($M_{post}$), (3) actions that the user has to perform online ($A_{online}$), and (4) actions that a user has to perform offline ($A_{offline}$).
We differentiate between emails because we interpret access to users data as the primary goal of the request. 
However, there might still be some open questions (\eg if profiling is performed) that were not answered by the time the data was shared. 
An example of an action that a user must perform online is that the user has to enter additional data in an online form (\eg legal name). 
On the contrary, scanning the user's official identification document (\eg passport) is a typical example of task a user has to perform offline.
We created our ``work load score'' as is because we wanted to penalize (1) if companies set up obstacles, (2) if companies ask for additional information, and (3) the more users have to interact with the company.
We adjusted the weights prior to our evaluation based on the process (\eg a priori we did not expect that we would have to sign affidavits and thus we adjusted the weight of ``offline actions'').
In Section~\ref{sec:process} we describe the procedure of how users can access personal data (of the companies in our analysis corpus) in more detail.
We assigned different weights to the categories and computed a weighted sum of the number of occurrences of the different events (\ie $C_i = M_{pre} \cdot 5 +  M_{post} \cdot 2 + A_{online} \cdot 10 + A_{offline} \cdot 30$).
The result of the work load determination and comparison between inquired companies is given in Figure~\ref{fig:success}.
The figure shows a clustered version of the SAR results.
We computed the distance between all points of the same ``response status'' (\eg ``got access'') and clustered the points that are close to each other, according to their euclidean distance. 
For each cluster we computed a new point at the arithmetic mean position of the x-y coordinates of the original points. 
The size of each new point shows how many points were combined to this point (larger points cluster more points).
To increase readability further, we set the response date of all companies who replied more than five days after the second deadline to five days after the deadline.

By the time of the first deadline of the first round of inquires (July 20th), we got access to our personal data from 13 companies (36.1\%), seven companies (19.4\%) told us that they do not store any personal data, and one company (2.8\%) told us that they would not grant us access to our personal data because they could not verify our identity.
A reason that a company might not have any data stored could be that they only capture specific events that we did not trigger (\eg clicks on ads).
Another reason why the companies might not have any data is that we provided a wrong cookie ID. Note that this unlikely since we pulled the cookie values directly from the \texttt{openWPM} database. 
In total, four (11.1\%) companies did not respond once, and eleven other companies (30.6\%) were still processing our request by the time of the first deadline.
Thus, above 40\% of all inquires where not handled within the legal period of 30 days. 

In round two, we only got access to our data by eight companies (19.5\%), 13 companies told us that they do not store any data related to the cookie id (31.7\%), and one company told us that they do not grant us access to the data (2.4\%). 
Nine companies did not finish the process in the period (22.0\%) and five companies did not respond once (12.2\%).
Thus, around 34\% did not finish the process in the legal period of 30 days.
During our first and second measurement, the amount of companies finishing the SAR process within 30 days did not change significantly.
However, more companies did state that they do not store data associated to a cookie ID.
It is notable that companies who stated that if one does not have an user account on their website, they would not store any data related to a cookie ID, did not respond to our SAR request within the legal deadline, in round two. One of these companies replied with our second deadline stating that they do not store any data related to the cookie ID.

Until the second deadline four\slash three more companies finished the process.
After our second deadline (30 business days after our inquiry), two more companies shared the collected personal data with us.
As for round two only one more company, that also took so long in round one, shared the data.

Eight companies interpreted the start date of the process as the day on which they got all the administrative data they need to process the inquiry. In all cases it was virtually impossible for the user to know upfront that this data was needed since the companies only shared the needed documents via email and did not mention them in their privacy policies. (\ie one company replied after seven days and asked for a signed affidavit. After we provided the affidavit they told us, another 5 days later, that they would \emph{``start the process''} and reply within 30 days.). 

In total (after the second deadline of round one), only 21 of 36 companies (53.9\%) shared the collected data, or told us that they do not store any data, 15 of 36 (41.7\%) were still in the process (or did not respond once), and one company said that it would not share the data with us because they cannot properly identify us.
In round two 63.9\% granted us access or told us that they do not store any data, 33.3\% did not finish the process, and, similar to round one, one company declined to grant access since they could not identify us.  
In these numbers, we \emph{excluded} companies that told us to address a subsidiary \slash parent company with our inquiry. 

Figure~\ref{fig:success} shows that if companies granted access, we see that work load is often quite low (in both rounds).
In one case with higher work load, in round 1, a long email exchange (in total 13 emails---six sent by us) was needed to get access, the other cases required a copy of the ID and in one case a signed affidavit.
It is notable that the overall work load in round 2 lowered and companies usually wrapped up the process faster.
The reduction of work load is because, on the one hand, we did not have to report broken SAR forms and on the other hand companies set up less offline obstacles.

Especially during round one, we observed that companies who claimed not to store any data still require intense interaction before giving that information. 
Two of these companies required a signed affidavit and a photocopy of our ID. The third company, after a longer email conversation, asked to call the customer support to explain our case in more detail, still coming to the result that they do not store any data.
All three companies did not respond in round two.

\begin{figure}[ht]
    \caption{Comparison of the work load to get access to personal data companies stored about a user. The time axis states when we got access to our data or date when the last mail was received. The graph differs between companies that gave us access to personal data they stored ($\medblackdiamond$), companies that claimed that they did not store any personal data ($\medblacktriangleup$), companies that did not grant access (\Flatsteel), and companies that did not grant us access to our data ($\medblackcircle$).}
    \label{fig:success}
    \centering
    \includegraphics[width=1\textwidth]{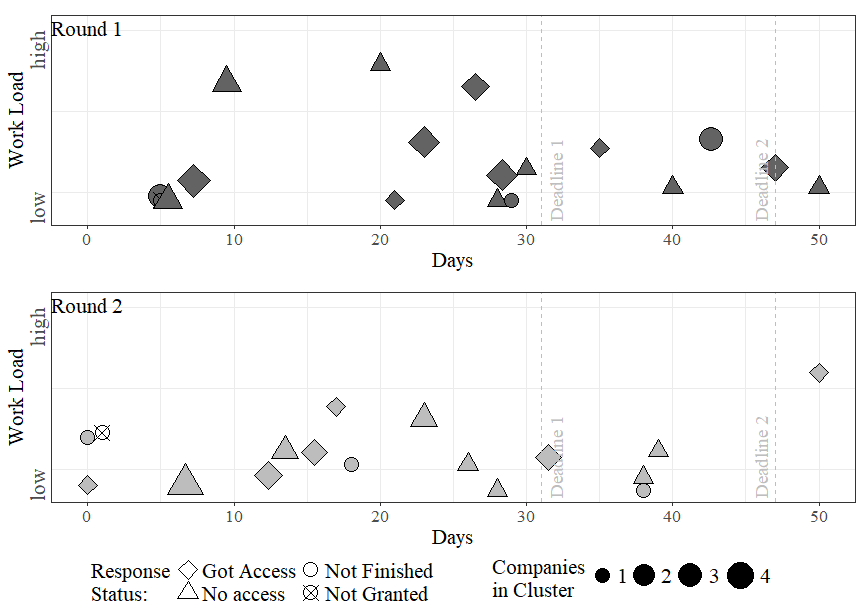}
\end{figure}

\subsection{Subject Access Request Process}
\label{sec:process} 
In this section, we describe the different SAR processes of the companies in our analysis corpus.
We found major differences how companies handle inquiries ranging from not responding at all, over simply sending the personal data via email, to sending (physical) letters which had to include a copy of a government-issued identification card and a signed affidavit, stating that the cookie and device belong to the recipient and only the recipient.

First nine, then ten companies presented us with additional obstacles to access personal data.
In total we had to sign four (round one)\slash three (round two) affidavits, send six\slash five copies of a government-issued identity card, and four\slash seven other forms of ``ID verification'' (\eg provide the used IP addresses).
26 \slash 25  companies did not set up any obstacles and four more companies did not reply once.

Most companies only require the user to provide the digital identifier (or directly read it from the browser's cookie storage) in order to grant access to the data associated to it.
Since most online forms do not provide all data a company collected about the user (\eg they provide the ad segments the company determined for the user but not the used IP addresses or visited websites) it is reasonable to grant access to this data if the cookie ID is provided.
However, online forms come with the risk that an adversary might fake the cookie ID and might get fraudulent access to personal data that is associated with another individual.
An affidavit is a way to counter this sort of misuse, and one company stated this as the reason for this additional step.

As described above, some companies require an affidavit or a copy of a government ID before they sent a copy of the data. The GDPR states companies ``\textit{should use all reasonable measures to verify the identity of a data subject who requests access}'', to make sure they do not disclose data to the wrong person.
Asking for identifying information is supposed to add a layer of security when data subjects request a copy of their data. The ad industry association emphasizes the possibility of this additional safeguard~\cite{iabeurope}, but official interpretations state that data processors should have ``\textit{reasonable doubts}'' before asking for additional data~\cite{article_29_portability}.
Those that request an ID did not explain their doubt and did not describe how the ID would help them to verify that the cookie ID is owned by the person requesting the data related to it. 

Access can mostly be requested via email or web form. Note that by the time we researched the ways to contact the companies, only six had a form to access the data.
During round one of our measurements, four more companies added such forms.
Since the responses of the forms did not answer all of our questions (\eg with whom the data is shared), we would have had to mail these companies anyway.
It is worth noticing that three of these forms did not work and we had to report the broken forms to the companies to use them.

Article 20 of the GDPR states that companies shall provide the data in a ``\emph{structured, commonly used and machine-readable format}''.
However, it is user-friendly if the data is provided in a ``human readable'' format (\eg a website) since it is easier to interpret than multiple lines of delimiter separated text.
Only one company provided both.
We categorized the received data into two categories: (1) 'machine', and (2) 'human' readable.
If the data is text based \emph{and} structured in a way that it can be parsed easily in an intuitive way (\eg delimiter separated values), we grouped it into category 1 (\eg \texttt{.csv} or \texttt{.txt} files). 
However, if the data was not in text-based (\eg a \texttt{.pdf} file) or could \emph{not} be parsed straightforward in a reusable way (\eg a Website that contains the information), we labeled it 'human readable'.

In total, we got seven \emph{machine} and six \emph{human} readable responses in round one.
As for round two, we got four \emph{machine} readable and four \emph{human} readable responses.
One company provided both in each round.
It appears that each type of access has a preferred way to return the data. 
Forms often return user-friendly output (\eg on a website), while data return via mail is exclusively in text format (\eg \texttt{.csv} files).

We also analyzed the positions of the employees who responded to our requests by analyzing their email signature, if possible.
Note that these numbers did not change for our two rounds---only that fewer companies replied (to improve readability we only use numbers from round 1).
Most requests (11) were answered by an anonymous person from the companies ``privacy team'', while eight were handled by the general support team of the company.
According to the job description of the person handling our request, four were handled by a person with a legal background (\eg a General Counsel or Paralegal).
In one case, a person neither from support nor legal team replied (\eg a Data Analyst), in three cases we only got automatic responses, and in five cases we could not derive the profession of the person (\eg they signed the mails with a name).

Even though several companies did not grant us access to our personal data, only one of these companies informed us (in both rounds) that they would extend the legal period which they are allowed to do, according to the GDPR.

Our data requests were not covered by the standard business process of some rather large companies (in terms of annual revenue) and due to this fact was not handled at all or handled unsatisfactorily. Some companies do not have a defined process for someone \emph{without} an account of their services who has a privacy-related question, although all companies are engaged in the collection and sharing of (unregistered) users they track online.
From a technical point of view, it is hard to argue to use cookies, which include personal identifiers, while not storing any data associated with it.
Furthermore, the privacy statements of these companies \emph{did} state that they use cookies to collect data about the users' online activity to perform targeted advertising.
In round two, these companies did not respond to our inquires.

\subsubsection{Answers to Our Questions}
Finally, we want to discuss the answers to the four questions we asked in the inquiries (see Section~\ref{sec:sar}).
Only few companies did answer specifically to the questions we asked. Most of them referred to their privacy policy or did not give any details at all. 
In round two, we got even less concrete answers than in round one.
In round one, overall 21 responses contained answers to Q1 (data types), six to Q2 (sources), nine for Q3 (profiling), and seven to Q4 (sharing). 
In round two, 23 companies replayed to Q1, six to Q2 and Q3, and four to Q4. 
Note that companies were not obliged to answer the question and that we could not check if the companies answered truthfully---if there is no public information in \eg the privacy statements that say otherwise (see also Section~\ref{sec:limitaions}).
With respect to Q1 and Q2, most answers contained references to or parts of the  privacy policy.
In the cases (in total three in each round) when companies told us with whom they exactly shared our data (\ie not a general list of partners), we could find these as edges in our graph.
Nine companies stated a general list of partners in their privacy policy.
For one of these companies we found two unlisted partners.

As Table~\ref{tab:policies} (Appendix~\ref{app:pp_overview}) shows, only a few companies (nine \slash seven) disclose whether or not they perform profiling there. Only one of the answers, where the privacy policy was unspecific, clearly stated that the data is not used for profiling.
Six answers described in some more detail how the data is processed, and would suffice the GDPR rule that ``\textit{meaningful information about the logic involved}'' should be provided, for example, one company explained that they make ``\textit{automated decisions about how much to pay for an ad placement and which ad placements to buy on behalf of advertiser clients.
The result of these decisions is merely that some fraction of the ads shown on your browser during your normal browsing activities will have been selected by X}''.
Only one company that stated that they do use profiling did not give any more details, and one other company stated in their email that they do not perform profiling, although their privacy policy says they do. 
In both cases, clear statements are necessary about whether or not they profiling processes could have any legal effects.

\subsubsection{Validation of Disclosed Information}
We also analyze if the information about data sharing were actually reflected in our measurement. Unfortunately, very few companies listed the actual partners with whom they specifically shared our personal data but only provided a general list of partners. 
We could find all specifically named sharing partners in our graphs. 
For one company, we found three additional partners that were not named in the privacy policy.
If a company shared clickstream data (three in total), we manually checked if we actually visited the websites and if the company did not include some websites that included that company.
We did not find a case in which a site was missing or another site was added, that we did not visit. 

The shared data is extremely heterogeneous in format (\eg \texttt{.pdf}, \texttt{.csv}, \texttt{.htm}, etc.), data contained (\eg segments assigned to the profile, clickstream data, IP addresses, etc.), and explanation of the data (\eg one company shared an \texttt{.csv} file with headers named $c_1$ to $c_{36}$ (sic.), one company had detailed explanations in an appended document, or another company told us that we should contact them if we had trouble understanding the data).

In terms of clarity\slash understandability of the provided data we also found various different approaches.
Some companies shared straight forward segments they inference from our (artificially) browsing behaviour (\eg Segment: \emph{Parenting - Millennial Mom}), others shared cryptic strings without explanation (\eg \emph{Company\_Usersync\_Global}), or data that was incorrectly formatted somewhere in the process (\eg \emph{your\_hashed\_ip\_address: Ubuntu} (sic.))
However, we did not find any instance where data was shared that was not mentioned to be collected in the privacy policy and many instances where not all collected data was shared.

The described usability\slash transparency problems and solutions how to fix them should be addressed in future work.
\section{Discussion}
\label{sec:discussion}

Our results show that tracking and sharing of information was reduced, but that there are obstacles for users during the SARs. We also discuss limitations and ethical considerations of our work. Finally, we describe possibilities for future work in this area.

\subsection{Assessment of the SAR Process}
The GDPR puts companies in quite a dilemma when it comes to the usage of cookies. 
On the one hand, they want to make it as easy as possible for users to exercise their right to access their personal data (\eg by providing a website that automatically displays all information based on the cookie set in the browser). On the other hand, they want to prevent misuse as some companies might leak long browsing histories of some users since most keep the data for over a year.

There is a general mismatch between the definition of personal data about one data subject and the socio-technical context. For example, it is not able to model use cases in which a device is used by multiple persons. It is impossible for these services to check who used a device and left the possibility that someone gets access to data that another user of the same device might have intentionally deleted.

\subsection{Limitations}
\label{sec:limitaions}

Our measured third-party graphs are only a small subset of the real third-party relations of a website. 
A website, even though we tried to mask it, might detect that a crawler is visiting the website and embed different objects or none at all.
Aside from randomly scrolling, we do not interact with the websites which might also influence our results because some third parties might only be embedded in a user performs a specific task (\eg if the user starts a purchase process, a third-party might be embedded to handle the credit card payment).
In this sense our dataset is biased.
However, we did not aim to capture all third parties, which is probably impossible in an automatic fashion, but wanted to asses the user transparency regarding the collected data of third parties.
In our opinion, the dataset is valid for that purpose because it covers the third parties that are used whenever a user visits a website which is a common use case of a website.

We only contacted 39 companies, which is a subset of all online advertising companies.
However, we have shown that the contacted companies come from different market areas and that they represent the most prominent companies (in our measurement).
Future work should focus on the usability of SARs in a user study and include more companies.

We cannot check whether or not the companies answered truthfully and provided all data they stored, shared, and processed.
To check that we would need to direct access to the services' databases.
For the same reason, we were also not able to measure what information companies exchanges on separate channels besides the synced IDs.

We do not measure the actual amount of data shared, but measure the amount of sharing connections of companies. 
To measure the actual data shared by the companies one would need access to the backend of the sharing companies.

\subsection{Ethical Considerations}
\label{sec:ethics}
Since our research includes human subjects (the persons exercising their rights and the persons responding to our requests), ethical considerations need to be taken into account.
In this work, we analyze the SAR process of different companies and not the persons replying in detail. 
Hence, we do not see any particular reason why we have to disclose that we conduct this survey.
We did not choose to debrief most of the companies since we exercise our right granted by the GDPR and we do not study the persons replying, but the process of how companies handle the new GDPR regulation.
Note that after our second deadline (in our first measurement) we contacted the companies that did not respond at all or had a poorly designed process, without any responses.

The data collection in this work is automated (\ie we use \texttt{openWPM}).
Thus, one might argue that these data are not related to a person.
However, we run the collection on a computer that is exclusively used by members of our team who also contacted the companies (i.e., we did not get access to any personal data that might be related to another person and that the data is to some extent related to this person).
Ultimately, we could have used the cookies of this person, but that would create problems when sharing our data since the data would include non-artificial personal data. 
When contacting the companies, we did not disclose we conduct a scientific survey, but we did disclose the real names of two authors in each mail and on the photocopied IDs.
We also answered all of the companies questions truthfully (\eg if we had been in contact with a company in any other way aside from this survey) and reported all problems (\eg data access forms that did not work) that we noticed during the process.
Furthermore, none of the emails were sent out automatically.

\subsection{Future Work}
The GDPR is entirely new, and therefore there are still several open research questions, from a cybersecurity point of view, that should be addressed in future work.
Some examples of such work are briefly discussed in the following.

\emph{Usability of  SARs}: As just mentioned we only contacted 39 companies. Future work should focus on the usability of SARs in a user study. In the study users should try to get access to their personal data of further companies which helps to understand how to improve the process.

\emph{Adversary  SARs}: As mentioned in this paper some companies do not validate the identity of the data subject but merely read the browser's cookie value. 
An open question is if an adversary could abuse this by faking the cookie value to get access to personal information of other individuals.

\emph{Privacy By Design}: Art. 25 of the GDPR ("\textit{Data protection by design and by default}") requires companies (on a technical and organizational level) to implement data-protection principles both at the time of data collection and processing.
Future Work could provide an overview of how companies realize such measures (\ie how do companies implement such measures) and check if there is a distinct difference of collected data before/after a user gives consent. 

\emph{Opting-out and Data Erasure}: In this work, we have demonstrated the ID sharing connections of websites.

Future Work could measure if opting-out has any effect on this (\eg user $A$ opts out at service $S_1$. Does sharing between $S_1$ and other services $S_2$ stop?), and if data erasure propagates through the network (\eg if a user asks $S_1$ to delete his\slash her personal data will the sharing partners of $S_1$ delete them as well?).

\section{Related Work}
\label{sec:related_work}

Generally speaking, previous work deals with online privacy measurements, online tracking, analyses of ad networks, and work on privacy statements.

\subsection{Online Privacy Measurements}
Most previous work analyzes online privacy through measurements --- which have all been conducted prior to the GDPR. Gonzales \etAl presented a large-scale study on the use of HTTP cookies~\cite{gonzalez2017cookie}.
The authors analyzed more than 5.6 billion HTTP requests over a period of 2.5 months.
They show that, in practice, cookies are much more sophisticated than simple \texttt{name=value} pairs and present an algorithm capable of inferring the format of a cookie with high recall and precision rates. 

In 2016, Englehardt and Narayanan published their work on measuring online tracking~\cite{Englehardt2016}.
They introduce the open-source measurement tool \textsc{openWPM} which they used to crawl and analyze the top one million websites on the Internet. 
They analyzed cookie-based and fingerprint-based tracking among 13 other types of measurements.

Another large-scale measurement study was conducted by Acar \etAl in 2014~\cite{acar2014web}.
In this paper, the authors examined canvas fingerprinting, evercookies, and the use of cookie syncing.
According to their study, 5\% of the top 100k websites use canvas fingerprints to identify users.

Most recent are the papers by Papadopoulos \etAl\cite{papadopoulos2018cookie} and Karaj \etAl\cite{Karaj2018}.
Papadopoulos \etAl performed a study on cookie syncing on a one-year long dataset including browsing activity from 850 mobile devices.
According to their measurement, over 97\% of users are exposed to cookie syncing and an ID is shared with 3.5 companies on average.
Karaj \etAl monitored the online tracking landscape over a period of ten months of real users through a browser extension.
They try to illuminate the online tracking business and argue that more transparency and accountability is needed since users struggle to keep control of their data.
The authors plan to continue their work and make their data available.
Another work from 2016 by Yu \etAl\cite{yu2016tracking} proposed a method for users to identify unsafe data elements which have the potential to identify individuals uniquely.
Based on their measurements, 95\% of all websites embed objects that might be used to track users and show that only 22\% of the page loads do \emph{not} transmit privacy-related data.

The introduced related work measures the tracking capabilities and other privacy implications of websites.
We focus on the relations between websites, how hard is to access data a company gathered, and how companies adopted the GDPR which has not been discussed in the past.  

Franken \etAl\cite{Franken2018} discuss the effectiveness of the \emph{Same Origin Policy}.
They present a framework to test the implementation of the policy and show that protection mechanisms of modern browsers and extensions can be bypassed.

\subsection{Ad Networks}
Additional work has been conducted regarding ad networks. 
Falahrastegar \etAl investigate the connections between third parties focusing on ID sharing~\cite{falahrastegar2016tracking}.
They find that domains show more syncing activities when a user is logged out and group the sharing domains based on their content.

The work of Castelluccia \etAl is also related to advertising~\cite{betrayed2012}.
The authors introduce a method to filter targeted ads and infer the users' interests from them.
Their results indicate that an adversary is capable of reconstructing user profiles even if she has access to a limited amount of ads. 

Most recently, Bashir \etAl introduced a so-called \emph{Inclusion} graph that models the diffusion of online tracking through Real-time Bidding~\cite{bashir2018diffusion}.
They show that 52 advertisers or analytic companies obverse over 90\% of an average user's online clickstream.
The work differs from ours since we do not want to shed light on the connection of online advertising companies, but measure effects of the GDPR.

A method to identify server-side information flow in the ad economy is present by Bashir \etAl\cite{bashir2018retarget}.
To do so, they use re-targeted ads to reveal information flows.
Kim \etAl recently presented their work on an ad budget attack~\cite{kim2018adbudgetkiller}.
They present an attack on targeted advertisers that legally drain the advertiser's ad budget.

\subsection{Computer Law and Privacy Policies}
Aside from the presented more technical papers, our work is related to work that focuses on the legal aspects of the GDPR. Recently, Libert presented his work on an automated approach to auditing disclosure of third-party data collection in websites' privacy policies~\cite{libert2018automated}.
The work shows empirically that it is unmanageable for a person to read the privacy policies of the first and third parties.

Gjermundr{\o}d \etAl presented a GDPR-compliant framework that allows users to create a cryptographically verifiable snapshot of her data trail \cite{Gjermundrod2016}.

Van der Auwermeulen~\cite{VANDERAUWERMEULEN201757} and De Hert \etAl\cite{DEHERT2018193} discuss the right to data portability from a computer law point of view.
De Hert \etAl give a systematic interpretation of the new right and propose two approaches on how to interpret the legal term ``data provided'' in the GDPR.
The authors argue that a minimal approach, where only data directly given to the controller (\eg data entered into a form) can be seen as ``provided''.
They also describe a broad approach which also labels data observed by the controller (\eg browser fingerprints) as ``provided''.
The author proposes to adopt the extensive approach.

Van der Auwermeulen~\cite{VANDERAUWERMEULEN201757}, on the other hand, compares the \emph{European Competition Law} with the \emph{U.S. antitrust law} if they could be applied in the case of data portability.
The author concludes that the U.S. law is not favoring data portability while the European law might be used for such purpose.
Furthermore, the author acknowledges that with the GDPR, which was just released and not in effect by the time when the article was published, it is clear that data subjects have the right to access their personal data.
\section{Conclusion}
\label{sec:conclusion}

Granting users the right to regaining control over their personal data is at heart of the European General Data Protection Regulation.
In this work, we give an overview of the GDPR's effect on the online ad economy.
We show its effect on cookie syncing, compare the ``subject access request'' process of different companies, and give an overview of how specific companies are connected to each other.

As for the cookie syncing between third parties, we show that the activities shrink by about 40\% after the GDPR took effect.
However, the general structure of how the third parties are connected stays more or less the same.
This indicates that the GDPR did not revolutionize the ad ecosystem, but rather has a direct effect on the amount of information sharing in the ecosystem.

Our analysis of the SAR processes shows that while most companies offer easy ways to access the collected personal data, some companies put up several obstacles for users to access it.
The obstacles range from signed affidavits over providing additional information (\eg phone numbers) to copies of official ID documents.
In our view, especially requesting a copy of an official ID document is not proportional for the use case.
The different approaches how personal data might be accessed show the insecurities and different interpretation of the new law.
Looking into the response behavior, we see that over 58\% of the companies did not respond within the legal period of 30 days, but only one company extended the deadline by two more months.
We could fill official complaints at our local Data Protection Authority for these companies.

Finally, we measured how the third parties that perform cookie syncing are connected.
We show that these companies are mostly arranged in star typologies, and therefore we find nodes that have various direct sharing partners, while others only have one direct partner and many indirect partners.
From this, we draw that a website might risk that it shares personal data of its users to any third parties without knowing.
From the users' perspective, this is highly undesirable especially because the GDPR does not require companies to name their sharing partners.
In this case, the users have virtually no chance to keep control over their data. 
In our experiment, only three companies named us the specific companies they shared our data with, others named all of their sharing partners (sometimes hundreds of partners), and other did not disclose any names at all.

The most important tool for online services to inform users about their data practices are their privacy policies, but our results show that not all companies take the legal obligations seriously. Besides services that miss information required by the GDPR, we found the majority of privacy policies do not sufficiently explain what data is shared with whom.

\section*{Acknowledgment}
This work was partially supported by the Ministry of Culture and Science of the German State of North Rhine-Westphalia (MKW grant 005-1703-0021 ``MEwM'' and ``NERD.NRW'') and the German Federal Ministry of Education and Research (BMBF grant 16KIS0395 ``secUnity'').

{\normalsize \bibliographystyle{acm}
\bibliography{ad_network_paper}}

\begin{appendices}
\section{Mail Template}
\label{app:mail}
In round one we used the following template to get in touch with the companies.
During round two we used similar sentences that contained the same information.
We replaced the \textbf{bold} text with information we extracted from our long term measurement.
We pulled the corresponding cookies for each domain from our measurement logs, and inserted the cookie IDs into the mails. 
As for the cookie syncing we extracted up to five, chosen randomly if necessary, instances of such syncing and inserted them into the mail.

\begin{framed}
\begin{flushleft}
\emph{Subject:}\linebreak Request for personal data and additional information

\emph{Body:}\linebreak
Hi,\linebreak
I noticed that there is a cookie stored in my browser associated with the  domains \textbf{Domain 1} and  \textbf{Domain 2} which are owned by your company. 
Citing my European privacy rights, I would like to ask you to answer the following questions:

1) What information about me\slash associated with that cookie do you store and process?

2) Where did you get that information from? Did you get it from third parties?

3) Do you use the data to perform profiling?

I'd also like to kindly request a copy of the data.\linebreak
The following cookies stored in my browser are associated with domains I found to be associated with your company:

 On domain: \textbf{Domain 1} key: \textbf{'id\_key\_1'} and value: \textbf{personal\_identifier\_1}
 
 On domain: \textbf{Domain 2} key: \textbf{'id\_key\_2'} and value: \textbf{personal\_identifier\_2}\linebreak
 
Another question I have is:\linebreak
4) With whom do you share what information and how?
 For example I saw that you used the following IDs with your partners: \newline
    partner: \textbf{Sync Partner 1} using the key: \textbf{'sync\_key\_1'} and value: \textbf{'sync\_id\_1'}\newline
    partner: \textbf{Sync Partner 1}  using the key: \textbf{'sync\_key\_2'} and value: \textbf{'sync\_id\_2'}\newline
    
Thanks for your support,\linebreak
First Name Last Name
\end{flushleft}
\end{framed}

\section{Graph characteristics}
\label{app:graphscharacteristics}
Table~\ref{tab:graphscharacteristics} presents general graph characteristics of our conducted measurements.
The longest possible distance between two nodes, modularity and medium degree of the graphs remains more or less stable. 
Nevertheless, the number of communities is reduced from 68 in M\#1 to 48 communities in M\#2 and M\#3, and even 43 communities in M\#4.
The average distance between node pairs in the graph indicates the average path length. 
These values do not change much across the three graphs we measured after the GDPR took effect.
We draw from this that the GDPR did not revolutionize the ecosystem.
\begin{table}[!htb]
\caption{Characteristics of our graphs w\slash o isolated nodes.}
\centering
\label{tab:graphscharacteristics}
\resizebox{\textwidth}{!}{
\begin{tabular}{@{}c|llllll@{}}
\toprule
\multirow{3}{*}{CW}          & diameter & medium & modularity & avg.        & avg.   & comm. \\ 
          &          & degree &            & clustering  & path   &       \\
          &          &        &            & coefficient & length &       \\
\midrule
20 (M\#1) & 9        & 2.975  & 0.576      & 0.234       & 3.125  & 68    \\
21 (M\#2) & 8        & 2.619  & 0.609      & 0.177       & 3.098  & 48    \\
26 (M\#3) & 8        & 2.518  & 0.622      & 0.181       & 3.234  & 48    \\
28 (M\#4) & 9        & 2.432  & 0.699      & 0.154       & 3.349  & 43    \\
34 (M\#5) & 10       & 2.291  & 0.659      & 0.158       & 3.192  & 37    \\
38 (M\#6) & 10       & 2.314  & 0.717      & 0.074       & 3.931  & 26    \\
43 (M\#7) & 9        & 2.36   & 0.714      & 0.071       & 3.5    & 38    \\
\bottomrule
\end{tabular}}
\end{table}

\section{Analysis Corpus}
\label{app:corpus}
Table~\ref{tab:ana_corpus} shows the companies included in our analysis corpus. 
In two cases (\texttt{Turn} and \texttt{BidSwitch}) parent organizations replied to our inquiries, in one case a subsidiary replied (\texttt{FreeWheel)} instead of the inquired company, and in one case (\texttt{RTL Group}) we were told to address our inquiry to a subsidiary (\texttt{SpotX}). 
Thus, our final corpus consists of 39 companies.

\begin{table}[!htb]
\centering
\caption{The companies of our analysis corpus grouped by their respective business field(s). AppNexus ($\bigstar$) and Adform ($\clubsuit$) run two services and are therefore listed twice. SpotX is a subsidiary of the RTL Group ($\dagger$).}
\resizebox{\textwidth}{!}{
\label{tab:ana_corpus}
\begin{tabular}{@{}llll@{}}
\toprule
\multicolumn{4}{c}{\textbf{\textsc{Supply-Side Platforms}}}\\
Improve Digital & Smart AdServer  & AppNexus\textsuperscript{$\bigstar$} & Rubicon Project \\
Index Exchange  & Sovrn \\
\midrule

\multicolumn{4}{c}{\textbf{\textsc{Demand-Side Platforms}}}\\
TripleLift  & MediaMath & Adform\textsuperscript{$\clubsuit$} & AppNexus\textsuperscript{$\bigstar$}\\
OpenX & DataXu & \multicolumn{2}{l}{IponWeb (BidSwitch)}\\
Sizmek & \multicolumn{2}{l}{Amobee (Turn)} \\
\midrule

\multicolumn{4}{c}{\textbf{\textsc{Advertising Companies}}}\\
The Trade Desk & Sharethrough & NeuStar & Criteo\\
Acxiom & SpotX\textsuperscript{$\dagger$}  & Quantcast\\
\midrule%

\multicolumn{4}{c}{\textbf{\textsc{Data Management Platforms}}}\\
Lotame & Adform\textsuperscript{$\clubsuit$} & \multicolumn{2}{l}{Media Innovation Group}\\
Drawbridge \\
\midrule

\multicolumn{4}{c}{\textbf{\textsc{Further Companies}}} \\
Google & Verizon & \multicolumn{2}{l}{FreeWheel (Comcast)} \\
Oracle & Adobe & RTL Group\textsuperscript{$\dagger$}  &Microsoft  \\
comScore & Twitter & \multicolumn{2}{l}{Harris Insights \& Analytics} \\
Facebook & Amazon\\

\bottomrule
\end{tabular}}
\end{table}

    \section{Analysis Corpus Classification}
    \label{app:partners}
    Table~\ref{tab:partners} lists direct syncing partners of our analysis corpus during our four measuring points (\ie With how many partners did a third party sync cookie IDs).
    Furthermore, the table shows the types of the nodes. (\ie outer, balanced, center, and isolated).
    The rows 'Remaining Nodes' show the mean syncing relations from all third parties that are not present in our analysis corpus.\\
    For most third parties, the direct partners are reduced over the course of our measuring points.
    The biggest reduction is attributed to \texttt{Google} and the only exception is \texttt{Adform}. 
    Interestingly, the number of direct connections from M2 to M3 increases again in some cases, for example for \texttt{Criteo}.
    In principle, the behavior of indirect partners is comparable to the behavior of direct partners which is not surprising, as the partners of the partners are dependent on the direct partners.
    Meaning that if the direct partners of one node are reduced the indirect partners are also likely to be reduced.
    In general, this means that personal data of users are less likely to be shared unnoticed with multiple parties.
    
    \begin{table}[!htb]
    \centering
    \caption{Synchronization relations of our analysis corpus. 'Direct Partners' indicates the amount of direct ID syncing. The types of the nodes outer (o), balanced (b),  center (c), and isolated (iso) are displayed as well.}
    \label{tab:partners}
    \resizebox{\textwidth}{!}{
    \begin{tabular}{@{}rl|rrrr|cccc|@{}}
    \toprule
    \multicolumn{10}{c}{\textsc{Analysis Corpus}}\\
    \# & \multicolumn{1}{c}{Third Party}    & \multicolumn{4}{c|}{Direct Partners}  & \multicolumn{4}{c|}{Type}\\ 
    && M\#1 & M\#2 & M\#3 & M\#4 & M\#1 & M\#2 & M\#3 & M\#4\\ 
    \midrule
        1. & Google         & 195 & 138 & 118 & 112 & c & c & c & c \\
        2. & Facebook       & 11  & 11  & 9   & 9   & c & c & c & c \\
        3. & Amazon         & 31  & 19  & 17  & 1   & c & c & c & b \\ 
        4. & Verizon        & 18  & 10  & 10  & 6   & c & c & c & c \\
        5. & AppNexus       & 69  & 42  & 40  & 44  & o & c & c & c \\
        6. & Oracle         & 30  & 31  & 27  & 18  & c & c & c & b \\
        7. & Adobe          & 11  & 8   & 5   & 4   & c & c & b & b \\
        8. & Smart AdServer & 1 & 1 & \multicolumn{2}{c|}{isolated}  & o & b & \multicolumn{2}{c|}{isolated} \\
        9. & RTL Group      & 16  & 8 & 7& 10 &c& c & c& c \\
        10. & Improve Digital & 2 & 1 & 1 & iso  & b & b & o & iso \\
        11. & MediaMath       & 16  & 7 & 8 & 10  & c & c & c& c  \\
        12. & TripleLift      & 5 & 1 & 2& 4  & b& o   & b & b       \\
        13. & RubiconProject  & 12 &\multicolumn{3}{c}{isolated} &  c &\multicolumn{3}{c|}{isolated}       \\
        14. & The Trade Desk  & 12 & 7& 5 & 6& c & c & b & c \\
        15. & ShareThrough    & 2 &\multicolumn{3}{c}{isolated}  &  b &\multicolumn{3}{c|}{isolated}    \\
        16. & Neustar         & \multicolumn{3}{c}{isolated} & 10  & \multicolumn{3}{c}{isolated}  & c      \\
        17. & Drawbridge      & 1 & iso & 1 & iso & o & iso  &e  & iso    \\
        18. & Adform         &1 & 14 &11 & 13 &o &c&c & c \\
        19. & Bidswitch      &3 & 5 & 3 & 2 & b & b & b& b \\
        20. & Harris Insights \& Analytics & 4& 2& 2 & 2 & b & b & b & b  \\
        21. & Acxiom         & 12 & 2 & 6& 5   &c & b & c & b \\
        22. & Index Exchange & 6  & 4 & 3 & 2 & c & b & b & b \\
        23. & Criteo         & 6  & 1 & 4 & 2 & c & o & b & b \\
        24. & OpenX          & 16 & 7 & 6 & 4 & c & b & b & e \\
        25. & DataXu         & 7  & 6 & 4 & 3 & b & c & b & b \\
        26. & Lotame         &2 & 3 & 3& 2    & o & b & b & b \\
        27. & FreeWheel      & \multicolumn{4}{c|}{isolated} & \multicolumn{4}{c|}{isolated} \\
        28. & Amobee         & \multicolumn{4}{c|}{isolated}   &\multicolumn{4}{c|}{isolated} \\
        29. & comScore       & 25  & 20 & 20& 17  & c& c & c& b \\
        30. & spotX          & \multicolumn{4}{c|}{isolated}   & \multicolumn{4}{c|}{isolated}   \\
        31. & Sovrn          & \multicolumn{3}{c}{isolated} & 2   & \multicolumn{3}{c}{isolated} & b        \\ 
        32. & Sizmek         & 23  & 14 & 18& 2 &c& c& c & b \\
        33. & Twitter        & \multicolumn{2}{c}{isolated} &2 & iso   & \multicolumn{2}{c}{isolated}  &o    & iso  \\
        34. & Microsoft       & iso & 1 & iso & iso  & iso  & o & iso  & iso \\
        35. & Media Innovation Group & 2   & 1 & 2& 2   & b &  o & o& o \\
        36. & Comcast        &5 & 4& 4 & 4  & b & b & b& o  \\
        37. & Turn    &2  & 2   & 2& 4 &  b&  b & b& b \\
        38. & Quantcast    & 2& 2  & 1 & 1 &b & b & o& o  \\
        39. & IponWeb        & iso   & 31 &\multicolumn{2}{c}{isolated}   & iso& c& \multicolumn{2}{c}{isolated} \\
    \toprule
    \multicolumn{10}{c|}{\textsc{Remaining Nodes}}\\
    &    & \multicolumn{4}{c|}{Mean Direct Partners} & \multicolumn{4}{c|}{Nodes}\\ 
    &Node Type&  M\#1 & M\#2 & M\#3& M\#4 & M\#1 & M\#2 & M\#3 & M\#4\\
    \midrule
        & \multicolumn{1}{l|}{Outer corners (o)}& 1.31 & 1.26  & 1.22 & 1.33  & 268 & 183 & 164 & 190 \\
        & \multicolumn{1}{l|}{Center nodes (c)} &12.56  & 10 & 9.33 & 7.4  &25& 8 & 6  & 5    \\
        & \multicolumn{1}{l|}{Balanced (b)}  &2.11 & 2.13& 2.09  & 2.21  & 205& 135 &127 & 106  \\
    \bottomrule
    \end{tabular}}
    \end{table}

    \section{Graphs}
    \label{app:graph}
    Figure~\ref{fig:measurment_graphs} displays all third parties (nodes) and their cookie syncing relations edges) observed during four of our conducted measurements (before (M\#1), right after (M\#2), one month after M\#3), and two months after (M\#4) the GDPR took effect). 
    The darker a node is, the more important it is according to the PageRank algorithm. 
    The edges are also coloured according to their importance. The darker the edge, the more important it is.
    A comparison of the graphs reveals a reduction from M\#1 to M\#2 and to M\#3 as to M\#4.
    \begin{figure}[ht]
            \centering
            \begin{subfigure}[b]{0.473\textwidth}
               \centering
                \includegraphics[width=\textwidth]{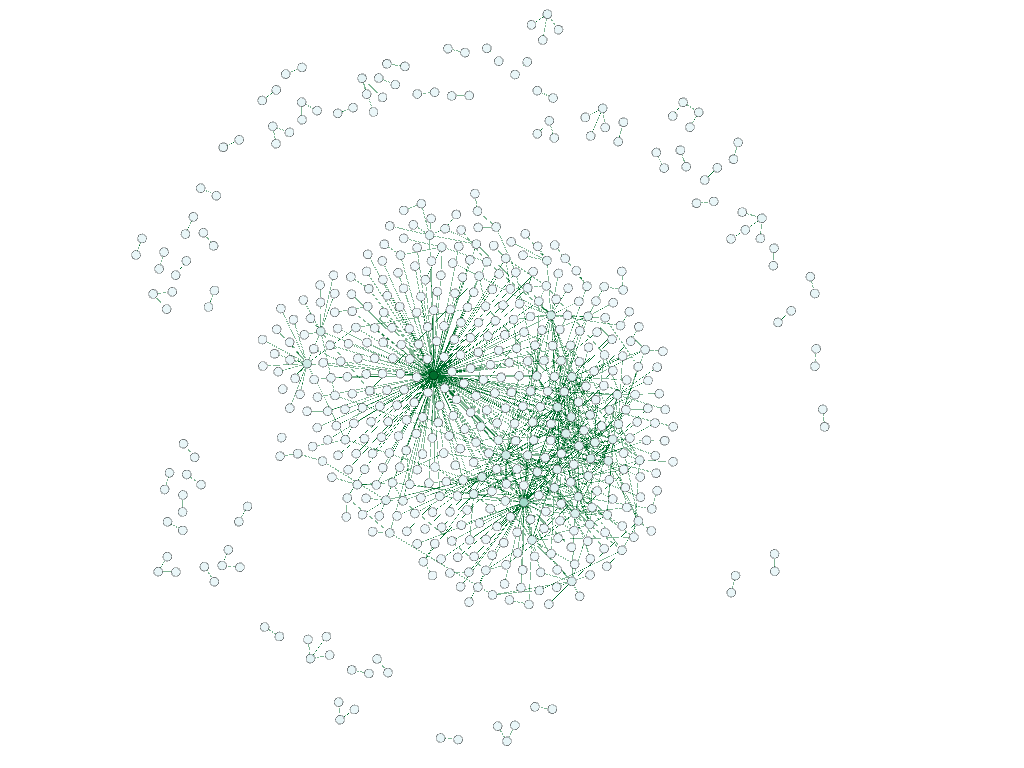}
                \caption{M\#1}    
            \end{subfigure}
            \hfill
            \begin{subfigure}[b]{0.473\textwidth}
                \centering 
                \includegraphics[width=\textwidth]{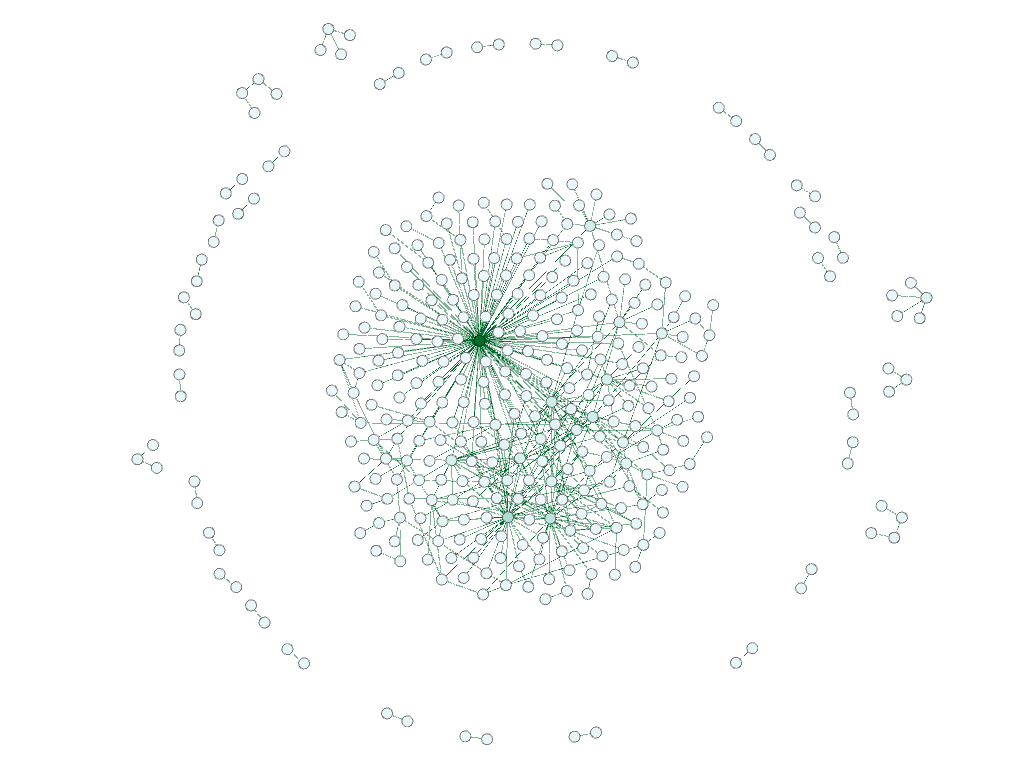}
               \caption{M\#2}   
            \end{subfigure}
        \vskip\baselineskip
            \begin{subfigure}[b]{0.473\textwidth}   
                \centering 
            \includegraphics[width=\textwidth]{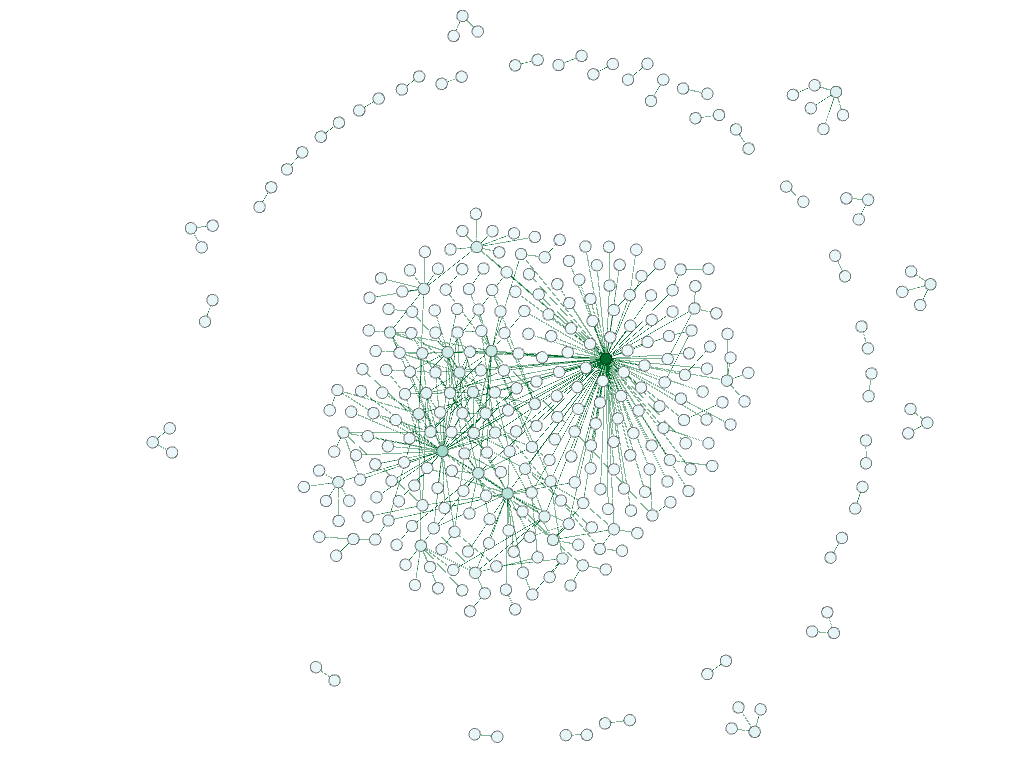}
                \caption{M\#3} 
            \end{subfigure}
            \quad
            \begin{subfigure}[b]{0.473\textwidth}   
                \centering 
                \includegraphics[width=\textwidth]{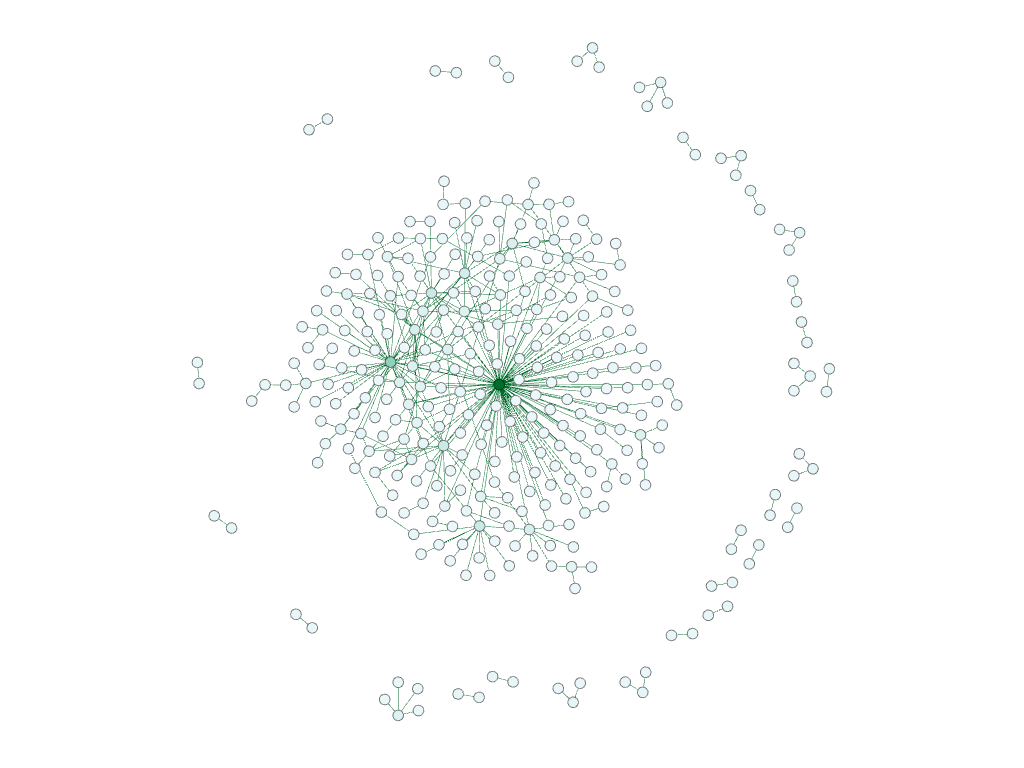}
                \caption{M\#4}
            \end{subfigure}
            \caption{Graphs of our measurements w\slash o isolated nodes.}
           
            \label{fig:measurment_graphs}
    \end{figure}

\section{Privacy Policy Overview}
\label{app:pp_overview}
Table \ref{tab:policies} provides a summary of the privacy policies of the companies in our data set. It lists the most important tracking and GDPR-related attributes and what information are disclosed.

\begin{sidewaystable*}[ht]
\centering
\resizebox{\columnwidth}{!}{
\captionof{table}{Overview of information available in privacy policies. * marks information that is required by the GDPR. \texttt{Legal Basis} refers to the sections in Article 6 of the GDPR: (a) consent, (b) contract, (c) legal obligation, (e) public, (f) legitimate interest; n.m. = not mentioned}
\label{tab:policies}
\begin{tabular}{@{}rlllllllllll@{}}
\toprule
 Company    &  Legal Basis*      & Shared Data  & 3rd CO* & Sensitive Data    & Profiling & Retention* & Partners*   & Data Access*   & DNT & Version \\ 
\midrule
 \href{https://policies.google.com/privacy?hl=de}{Google}     
            & a,b,c,f          & unspecified                 & y             & n.m.             & n.m       & unspecified & 7            & account       & n.m. & 05/2018\\
 \href{https://www.facebook.com/privacy/explanation}{Facebook}    &  a,b,c,d,e,f      & unspecified  & y & y    & n.m. & differs & categories  & account   & n.m. & 04/2018\\
 \href{https://www.amazon.com/gp/help/customer/display.html/?nodeId=468496}{Amazon}    &  n.m.      & unspecified  & n.m. & n.m.    & n.m. & n.m. & categories  & n.m.   & n.m. & 08/2017 \\ 
 \href{https://www.verizon.com/about/privacy/international-policy}{Verizon}    
            & a,b,c,f           & unspecified                 & y             & n.m.             & n.m       & unspecified & 329            & website, email         & n.m. & 05/2018\\
 \href{https://www.appnexus.com/en/company/platform-privacy-policy}{AppNexus}   
            & a,f               & unspecified            &y & n.m.             & n.m       & 3-60d, up to 18m         & 2309    & website & n.m. & 05/2018\\
 \href{https://www.oracle.com/legal/privacy/marketing-cloud-data-cloud-privacy-policy.html}{Oracle}     
            & a,c,f             & unspecified            & y             & \href{http://www.bluekai.com/health-related-categories.pdf}{health related}             & n.m.      & 12-18m & categories      & website & y & 05/2018\\
 \href{https://www.adobe.com/privacy/marketing.html\#online-advertising}{Adobe}     
            & a,b,c,f           & unspecified           & y             & n.m.             & y          & until opt-out  & categories & email, form     & n &  05/2018 \\
 \href{http://smartadserver.com/company/privacy-policy/}{Smart AdServer}
             & a,f             & unspecified & y             & n.m.             & y          & 1d-13m   & categories  &  email       & n.m. & 05/2018\\
 \href{http://www.rtlgroup.com/en/special/terms_of_use.cfm}{RTL Group}
             &  a,c,f           & unspecified  & y          & n.m.             & n.m.       & as long as necessary  & categories  & email        & n.m. & unclear\\
 \href{https://www.improvedigital.com/platform-privacy-policy/}{Improve Digital}
             & a                 & listed 

                                                                        & y             & n.m.             & n.m.          & 90d          & categories     & email        & y & 05/2018 \\
 \href{http://www.mediamath.com/privacy-policy/}{MediaMath}

             & f                 & unspecified        & y            & health related & y          & up to 13m     & categories &email     & n.m. & 05/2018\\
 \href{https://triplelift.com/privacy/}{Triplelift}

             & a,f                & unspecified        & y         &  ask to avoid  & n.m          & as long as necessary & categories & website              & n  & 05/2018\\
 \href{https://rubiconproject.com/privacy-policy/}{RubiconProject}

             & a,b,c,f            & unspecified        & y & n.m.    & n.m. & 90-366d & categories & form   & n.m. & 05/2018\\ 
 \href{https://www.thetradedesk.com/general/privacy}{The Trade Desk}

            &  a,f      & unspecified        & US & not allowed    & n.m. & 18m, 3y aggregated & categories  & \href{https://www.adsrvr.org/}{website}   & n.m. & 10/2018 \\ 
 \href{https://platform-cdn.sharethrough.com/privacy-policy}{ShareThrough}

            &  a,b,c,f         & unspecified        & y             & n.m.          & y & 13m & categories  & email   & n.m. & 05/2018\\ 
 \href{https://www.home.neustar/privacy/privacy-policy}{Neustar}

            &  n.m.     & IDs, segments       & US & not allowed    & n & 13m + 18m aggregated & categories & email   & n.m. & 08/2018\\ 
 \href{https://www.drawbridge.com/privacy}{Drawbridge}

             &  n.m.      & IDs, segments         & US & \href{http://www.bluekai.com/health-related-categories.pdf}{health related}    & n.m. & n.m. & categories & email   & n & 08/2018\\ 
 \href{https://site.adform.com/privacy-center/platform-privacy/}{Adform}

            &  a,f      & unspecified        & y & not allowed    & n.m. & 13m & \href{https://site.adform.com/privacy-center/adform-cookies/}{33} & form/email   & n.m. & unclear    \\ 
\href{http://www.bidswitch.com/privacy-policy/}{Bidswitch}

            &  a,b,c,f      & unspecified        & y & n.m.    & n.m. & ``as long as necessary'' & categories & n.m.   & n.m.  & 05/2018\\ 
 \href{https://www.visualdna.com/privacy-policy/}{Harris I \& A}\footnote{Harris Insight is the parent company, but for the analysis we reffered to the privacy policy of their online advertising subsidiary visualDNA}

            &  a,c       & listed        & y & y    & n.m. & purpose fulfilled & categories  & email   & n.m. & 07/2018 \\ 
 \href{https://www.acxiom.com/about-us/privacy/gdpr/}{Acxiom}
            &  a,f       & categories        & y & no    & n.m. & unspecified & categories & register   & n.m. & 05/2018 \\ 
 \href{http://www.indexexchange.com/privacy/}{IndexExchange}

            &  n.m.      & aggregated only        & US & no    & no  & 13m & categories  & \href{http://sar.indexexchange.com/}{website}   & n & 09/2018\\ 
 \href{https://www.criteo.com/privacy/}{Criteo}

            &  a      & aggregated        & y & no    & n.m. & 13m   & \href{https://www.criteo.com/privacy/criteo-works-with-the-following-platforms/}{61} & email/mail   & n & 05/2018\\ 
 \href{https://www.openx.com/legal/privacy-policy/}{OpenX}
            & a,f      & unspecified        & US & n.m.    & n.m & unspecified & categories  & email   & y  & 05/2018\\ 
 \href{https://www.dataxu.com/about-us/privacy/data-collection-platform/}{DataXU}

            &  a,b,c,f       & behavioural        & y & not in EU    & n.m. & 13m & categories & email   & n & 06/2018 \\ 
\href{http://www.lotame.com/about-lotame/privacy/lotames-products-services-privacy-policy/}{Lotame}    &  n.m.      & unspecified  & US & health related    & n.m & 13m & categories   & \href{https://www.lotame.com/about-lotame/privacy/privacy-manager-opt-out/}{website}   & y & 09/2018 \\ 
\href{http://freewheel.tv/privacy-policy/?noredirect}{FreeWheel}    &  a,b,f     & unspecified  & Y & n.m.    & n.m. & 18m & categories   & email   & n & 05/2018\\ 

\href{https://www.amobee.com/trust/gdpr}{Amobee}    

&  a,f      & unspecified  & US &  n.m.   & y & 13m & categories   & \href{https://www.amobee.com/gdpr/}{website}   & n.m. & 06/2018 \\ 
\href{https://www.comscore.com/ger/About-comScore/Privacy-Policy}{comScore}

&  a,b,c,f      & unspecified  & y & n.m.    & n.m. & n.m. & categories   & website   & n.m. & 12/2017 \\ 
\href{https://www.spotx.tv/privacy-policy/}{spotX}    

&  a,f      & listed  & n.m & n.m    & n.m. & 18m & 65 & website  & y & unclear \\ 
\href{https://www.sovrn.com/privacy-policy/}{Sovrn}

&  a,c,f      & n.m  & y & n.m.    & y & n.m. & unspecifc & \href{https://www.sovrn.com/yourdata/}{webform}   & n.m. & 05/2018 \\ 

\href{https://www.sizmek.com/privacy-policy/}{Sizmek}

&  a,b,c,f      & segments   & y & not knowingly    & n.m. & 13m & unspecified & \href{https://www.sizmek.com/rights-request/}{website}   & mixed & 05/2018\\ 
\href{https://twitter.com/privacy}{Twitter} 

&  \href{https://help.twitter.com/en/rules-and-policies/data-processing-legal-bases}{a,b,c,f}      & listed  & y & not allowed    & n.m. & 18m & \href{https://help.twitter.com/en/safety-and-security/data-through-partnerships}{16} & account   & n  & 05/2018\\ 
\href{https://privacy.microsoft.com/en-us/privacystatement}{Microsoft}    &  a,b,c,f      & unspecified  & y & y

& n.m, & 13m & \textgreater 9 & account  & n & 10/2018\\ 
\href{http://www.themig.com/en-us/privacy.html}{Media Innovation}    &  a      & unspecified  & US & n    & n.m. & 14m & partners & n.m.   & n.m.  & 09/2011\\
\href{https://www.quantcast.com/privacy/}{Quantcast}    &  a,f,      & listed  & y & not in EU    & n.m. & 13m & \href{https://www.quantcast.com/privacy/quantcast-partners/}{33}  & \href{https://www.quantcast.com/privacy/data-subject-rights/}{website}   & n.m. & 05/2018\\ 
\bottomrule
\end{tabular}

}
\end{sidewaystable*}
\end{appendices}

\end{document}